\documentclass[%
 reprint,
nofootinbib,
 amsmath,amssymb,
 aps,
prx,
]{revtex4-2}

\usepackage{mathptmx}
\usepackage{xcolor}
\usepackage{graphicx}
\usepackage{dcolumn}
\usepackage{bm}
\usepackage{amsfonts}
\usepackage{amsmath}
\usepackage[bottom]{footmisc}
\usepackage{hyperref}

\begin{document}

\preprint{APS/123-QED}

\title{\color{black}{Embedding stochastic dynamics in latent manifolds of recurrent neural networks}}

\author{Ramón Nartallo-Kaluarachchi$^{*}$}
\author{Renaud Lambiotte$^{\dagger,\ddagger}$}%
\author{Alain Goriely$^{\dagger}$}
\affiliation{%
$^*$Department of Applied Mathematics and Theoretical Physics, University of Cambridge, Wilberforce Road, Cambridge CB3 0WA, United Kingdom\\
$^{\dagger}$Mathematical Institute, University of Oxford, Woodstock Road, Oxford, OX2 6GG, United Kingdom
\\
$^\ddagger$Complexity Science Hub, Metternichgasse 8, Vienna, 1030, Austria\\
\text{{\normalfont\{rdn23\}@cam.ac.uk;\{renaud.lambiotte\}\{alain.goriely\}@maths.ox.ac.uk}}
}%

\date{\today}

\begin{abstract}
{\color{black} Recurrent neural networks (RNNs) provide a theoretical framework for understanding computation in biological neural circuits through their ability to support rich time-dependent behaviour. Whilst high-dimensional by construction, they can be used to encode low-dimensional dynamical systems, for example by restricting to low-rank connectivity, hence mirroring the emergence of latent neural manifolds found in empirical neural data. Here, we exploit a relationship between RNNs and a class of universal approximators to train networks that can represent arbitrary, nonlinear stochastic differential equations (SDEs), with given drift and diffusion coefficients, within a low-dimensional latent subspace. We show that RNNs can faithfully embed the drift and diffusion of a given SDE, including nonlinear and nonequilibrium dynamics such as chaotic attractors. We apply this approach by constructing RNN realisations of stochastic systems that transiently explore various attractors through both input-driven switching and autonomous transitions driven by nonequilibrium currents, which we interpret as models of associative and sequential (episodic) memory. To elucidate how these dynamics are encoded in the network, we introduce decompositions of the RNN based on its asymmetric connectivity and its time-irreversibility. Our results show that neural populations can implement a broad class of stochastic dynamics within low-dimensional manifolds, unifying ideas from the dynamics of neural computation, the theory of neural manifolds, associative memory, and nonequilibrium statistical mechanics.}
\end{abstract}

\maketitle


\section{Introduction}
\label{sec: intro}
Understanding how populations of neurons encode information and perform computations is one of the most significant problems in neuroscience \cite{Mathis2024decoding}. Central to the `connectionist' perspective is the idea that cognition is performed by distributed networks of interacting neurons. However, unlike the feed-forward architectures that dominate modern machine learning, the neuronal networks of the brain are recurrent, leading to a rich repertoire of time-evolving dynamics. As a result, classes of \textit{recurrent neural networks} (RNNs) have become the prototypical models for dynamic computation in biological neural circuits \cite{Vyas2020computation}. These networks, and variants thereof, have been used extensively to model neural data \cite{Pandarinath2018lfads, Durstewitz2023reconstructing}, develop theories of neural computation \cite{Sussillo2009chaotic, Hennequin2014optimal}, and perform sequence learning tasks, such as the prediction and encoding of chaotic time-series \cite{Gauthier2021nextgen, Kim2023neuralmachinecode,Kong2024reservoir}.

{\color{black} We consider an RNN composed of $n$ neurons}, each with internal state $u_i$ and output state $v_i = h(u_i)$, where $h$ is a monotonic nonlinear activation function (we will take $h(x) = \tanh(x)$). The dynamics of the internal state are given by the stochastic differential equation (SDE),
\begin{align}
    d\mathbf{u}(t) & = \mathbf{F}(\mathbf{u})\;dt + B \;d\mathbf{w}(t),\label{eq: RNN}
\end{align}
where $\mathbf{u} = (u_1,..,u_n)$ and $\mathbf{F} = (F_1,...,F_n)$ is the \textit{drift} function, with,
\begin{align}
    F_i & = -u_i + \sum_{j}W_{ij}v_j + I_i. 
\end{align}
Here $B \in \mathbb{R}^{n \times d}$ and $\mathbf{w}(t)$ is a Wiener process with $d$ independent sources of noise \cite{pavliotis2014stochproc}. We also define $\Upsilon = BB^{\top}/2$ to be the positive semidefinite \textit{diffusion matrix} of rank $d$. The connectivity from neuron $j$ to neuron $i$ is given by $W_{ij}$, and $I_i$ is the input current to neuron $i$.

\begin{figure*}
    \centering
    \includegraphics[width=\linewidth]{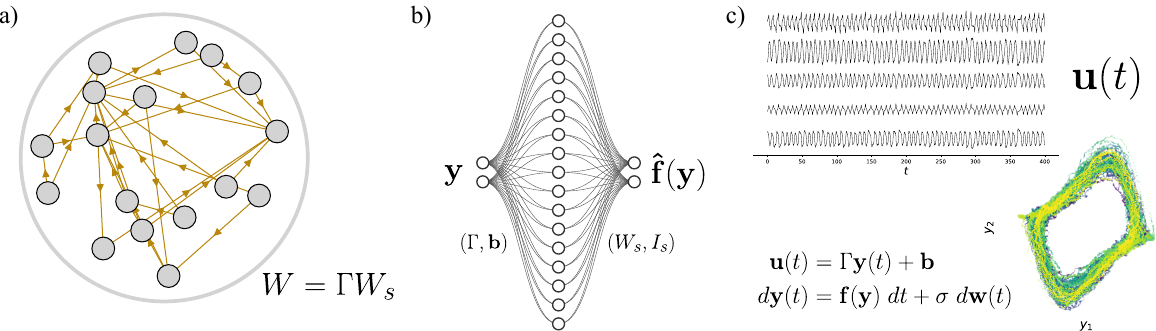}
    \caption{\textbf{Embedding dynamics in latent manifolds of RNNs.} $a)$ Our framework focuses on RNNs with low-rank connectivity of the form $W=\Gamma W_s$, where $\Gamma \in \mathbb{R}^{n \times k}$ with $k\ll n$. We train these autonomous systems to encode arbitrary target SDEs in a learnt affine subspace. $b)$ The training of the RNN can be reformulated as the training of a two-layer perceptron to approximate a target vector field. $c)$ Once the parameters of the RNN have been fitted, it can be integrated over time as an autonomous circuit which produces stochastic dynamics. Here we plot traces from 5 of 64 neurons in a network trained to approximate the van der Pol oscillator. When we project the dynamics into the learnt subspace, we recover a trajectory from the stochastic van der Pol system, which has been embedded into the RNN.}
    \label{fig: framework}
\end{figure*}

{\color{black} Encoding dynamical systems by approximating their vector fields with RNNs is a common method for investigating the dynamic repertoire of neural circuits \cite{Vyas2020computation}, such as the study of attractor neural networks \cite{Amit1989modelingbrain}. Common frameworks include the Neural Engineering Framework (NEF) \cite{Eliasmith2003neuralengineering}, reservoir computing through the `echo-state' framework \cite{Jaeger2001echostate}, or `FORCE' learning \cite{Sussillo2009chaotic}. Typically, one considers a low-dimensional `read-out' -- a projection -- of the high dimensional network activity i.e. $\mathbf{y}(t) = \phi(\mathbf{u}(t))$, which is considered to be the output that matches the target dynamical system \cite{Sussillo2013opening}.

The idea that high-dimensional neural circuits encode low-dimensional dynamical systems is rooted in the study of empirical neural recordings. Dimensionality reduction techniques have shown the emergence of low-dimensional, latent structure in high-dimensional neural activity associated with tasks like motor control, navigation, and working memory \cite{Perich2025neuralmanifold, Gallego2017manifolds}. These so-called \textit{neural manifolds} indicate that high-dimensional brain network activity is, in fact, constrained to low-dimensional subspaces where the dynamics are often mechanistically interpretable.

Low-dimensional dynamics in RNNs and the theory of neural manifolds have been linked through the study of \textit{low-rank} RNNs \cite{Mastrogiuseppe2018linking, Schmutz2025highdim, Arora2025efficient, Beiran2021shaping, Schuessler2020dynamics, Schuessler2022interplay, Valente2022extracting}, where a low-rank connectivity structure restricts the dynamics to latent subspaces. In particular, low-rank RNNs with linear read-outs can be associated with a class of universal approximators, which has allowed RNNs to be trained to represent arbitrary dynamical systems \cite{Eliasmith2003neuralengineering, Arora2025efficient, Beiran2021shaping}. In this article, we build on this direction by showing that RNNs with additive noise can encode target latent stochastic dynamics. We train low-rank RNNs to represent arbitrary nonlinear SDEs with additive noise in a latent affine subspace by directly matching the drift and diffusion of the target process using backpropagation of error (Fig.~\ref{fig: framework}). We link the encoding of latent stochastic dynamics to the emerging area of nonequilibrium statistical mechanics in neural computation and associative memory \cite{yan2013landscape,nartallokalu2025review,Behera2023active,Spisak2025self,Aguilera2025nonequilibriumassociative, Coolen2001statics, Coolen2001dynamics}.} In order to illustrate how the framework can be linked to associative memory, we introduce two classes of SDEs, both with multiple attracting states. The first is the gradient of a tiltable energy potential, which allows the network to choose between energy minima using input-driven switching. The second is a nonequilibrium stationary diffusion that cycles autonomously between attractors in a specified order using irreversible currents. {\color{black} We present these as illustrative models of both input-driven associative and sequential (episodic) associative memory \cite{Buonomano2023time}, showing that these processes can be encoded in a latent subspace of an RNN.} In order to investigate how the dynamics are encoded in the network, we introduce two decompositions of an RNN, the first based on the asymmetry of the connectivity, and the second on the time-irreversibility of the RNN. This theoretical model suggests that populations of neurons with asymmetric connectivity are capable of encoding a wide array of dynamical systems, including attractor-switching and -cycling dynamics, {\color{black} thus offering a mathematical extension for low-rank RNNs that links to the nonequilibrium and stochastic nature of neural recordings, and the theory of neural manifolds.}

\section{Recurrent neural networks with target latent dynamics}
\label{sec: target matching}

\subsection{Training the RNN}
\label{sec: ddm}

We focus on the RNN of the form,
\begin{align}
    d\mathbf{u}(t) & = \mathbf{F}(\mathbf{u})\;dt + B \;d\mathbf{w}(t),\\
    F_i & = -u_i + \sum_{j}W_{ij}v_j + I_i, \notag
\end{align}
for $i=1,...,n$, and consider a latent \textit{affine subspace},
\begin{align}
    \mathcal{A} = \{\mathbf{u} \in \mathbb{R}^n: \mathbf{u} =  \Gamma \mathbf{y} + \mathbf{b}, \mathbf{y}\in \mathbb{R}^k\},
\end{align}
for some $\Gamma \in \mathbb{R}^{n \times k}$ and $\mathbf{b} \in \mathbb{R}^n$ for the latent dimension $k\ll n$. We first ask if $\{W,I,B,\Gamma,\mathbf{b}\}$ can be chosen such that the dynamics of our latent parametrisation, $\mathbf{y}(t)$, follow a target SDE of the form,
\begin{align}
    d\mathbf{y}(t) & = \mathbf{f}(\mathbf{y})\;dt + \sigma \;d\mathbf{w}(t),
\end{align}
where $\mathbf{f}$ is the drift of the target process, and $\sigma \;d\mathbf{w}(t)$ is isotropic noise in $\mathbb{R}^k$.\footnote{{\color{black} Throughout, we will focus on encoding latent SDEs with isotropic diffusion. Nevertheless we note that this framework can be extended to encode processes with spatially-constant, but non-isotropic, noise structures. Given that the RNN has additive noise, we cannot encode processes in an affine subspace that have multiplicative noise.}}

{\color{black}The subspace $\mathcal{A}$ and the signal $\mathbf{y}(t)$ bear a resemblance to typical linear `read-outs' from RNNs. However, it is important to note that, in our setting, they represent a latent process over a neural manifold encoded within the high-dimensional activity of the network, with a focus on the case of associative memory. As a result, it is not appropriate to consider a `feedback signal' between the RNN output and its target, such as that required in a number of classical methods such as `echo-state' \cite{Jaeger2001echostate}, `FORCE' learning \cite{Sussillo2009chaotic}, or recursive least-squares \cite{Arora2025efficient}. These methods are applicable in generic frameworks such as reservoir computing and limited neuroscientific scenarios, such as motor control, but not in the study of neural manifolds for associative memory. The most established and general framework (albeit not biologically plausible) for training RNNs to perform sequence learning tasks is \textit{backpropagation through time} (BPTT) \cite{Werbos1990bptt}, which has been successfully applied to approximate dynamical systems with low-rank networks \cite{Dubreuil2022therole}.} Nevertheless, the training procedure is known to be computationally expensive, especially for stochastic data, and numerically unstable, often suffering from vanishing and exploding gradients \cite{Bengio1994learning}.

Additionally, a number of biologically-inspired methods have been developed to train biological neural networks. These include biologically-plausible methods for back-propagation \cite{Whittington2019theories}, plasticity rules for spiking neural networks \cite{Caporale2008spike}, Bayesian theories of neural computation \cite{Pecevski2011probabilistic}, feedback-based learning for models of motor control \cite{Hennequin2014optimal, Sussillo2009chaotic}, and local weight updates based on neural activity \cite{Bellec2020asolution, Murray2019local, Miconi2017biologically, Brunel1996hebbian}. The latter are often based on organising principles such as \textit{Hebbian learning} \cite{hebb1949hebbianlearning}.

{\color{black} Instead, we will show that training a low-rank RNN to approximate an SDE with additive noise is equivalent to training a two-layer perceptron to approximate the drift field, with an additional constraint for the diffusion term. {\color{black} This is not an example of sequence learning as the network approximates a vector field directly. A derivation of this equivalence for the deterministic case can be found in Ref.~\cite{Beiran2021shaping}, with relevant extensions in \cite{Arora2025efficient, Valente2022extracting}.} This equivalence has been previously exploited to train `read-out' layers from RNNs with quenched random connectivity with the aim of approximating {\color{black} an arbitrary vector field} e.g. via linear regression or backpropagation \cite{Eliasmith2003neuralengineering, Arora2025efficient}. Here, we will extend this approach to include additive correlated noise in the RNN which leads to isotropic diffusion in the latent process. Moreover, we will not focus on a random connectivity matrix, and will instead learn the weights, inputs, noise structure, and latent subspace via backpropagation of error. In App.~\ref{app: regression}, we show that backpropagation significantly outperforms ridge regression in this context. {\color{black} We note that our approach is distinct from sequence learning approaches, including BPTT.}}

Firstly, we assume that $\Gamma$ has full rank $k$, thus it has a Moore-Penrose (MP) inverse such that $\Gamma^{\dagger}\Gamma = \mathbf{I}_k$, where $\mathbf{I}_k$ is the identity in $\mathbb{R}^k$. Using Itô's rule \cite{pavliotis2014stochproc}, we have that the drift in the affine subspace is given by,
\begin{align}
    \hat{\mathbf{f}}(\mathbf{y}) & = \Gamma^{\dagger}\left(\mathbf{F}(\Gamma \mathbf{y} + \mathbf{b})\right),\label{eq: proj1}
\end{align}
whilst the diffusion matrix is,
\begin{align}
    \Sigma & = \Gamma^{\dagger}BB^{\top}(\Gamma^{\dagger})^{\top}.
\end{align}
However, matching $\hat{\mathbf{f}} = \mathbf{f}$ and $\Sigma = \sigma^2 \mathbf{I}_k$, is not sufficient to guarantee that projections of trajectories from the RNN, will behave as trajectories from the target SDE. This is due to the presence of the drift and diffusion terms in directions orthogonal to the affine subspace. To guarantee that the training procedure enforces that projections of the RNN trajectory recover trajectories from the target SDE, we must ensure there is a one-to-one relationship between the latent point $\mathbf{y}(t)$ and the high-dimensional $\mathbf{u}(t)$. Each point in the trajectory can be uniquely decomposed as,
\begin{align}
    \mathbf{u}(t) = \mathbf{u}_{||}(t) + \mathbf{u}_{\perp}(t),
\end{align}
where $\mathbf{u}_{||} = \Gamma\mathbf{y} + \mathbf{b} \in \mathcal{A}$, whilst $\mathbf{u}_{\perp} \perp\text{Im}(\Gamma)$ i.e. $\mathbf{u}_{\perp}$ is orthogonal to the image of $\Gamma$. As a result, $\Gamma^{\dagger}(\mathbf{u}-\mathbf{b}) = \mathbf{y}$ for all different $\mathbf{u}_{\perp} \perp \text{Im}(\Gamma)$, thus it is not injective.

To prevent dynamics in directions that are not `visible' under the projection, we enforce a low-rank parametrisation of the RNN,
\begin{align}
\label{eq: parametrisation}
    W = \Gamma W_s, && I = \Gamma I_s + \mathbf{b}, &&& B=\Gamma B_s,
\end{align}
where $W_s \in \mathbb{R}^{k \times n}$, $I_s \in \mathbb{R}^{k}$, $B_s \in \mathbb{R}^{k \times d}$, where we recall that $d$ is the number of independent noise sources. The result of this parametrisation is that, for any $\mathbf{u} \in \mathcal{A}$, we have that,
\begin{align}
    \mathbf{F}(\mathbf{u}) = -\mathbf{u} + \Gamma W_s  \mathbf{h}(\mathbf{u}) + \Gamma I_s + \mathbf{b} \in \text{Im}(\Gamma), 
\end{align}
where $\mathbf{h}(\mathbf{u}) = (h(u_1),...,h(u_n))$. Moreover, we have that $\Gamma B_s \;d\mathbf{w}(t) \in \text{Im}(\Gamma)$. Together, these imply that, given $\mathbf{u}(0)\in \mathcal{A}$, the drift and diffusion are tangent to the subspace, and thus $\mathbf{u}(t)\in \mathcal{A}$ for all $t>0$ i.e the affine subspace is \textit{invariant} under the dynamics. When the process begins outside the subspace, the `leak term', $-\mathbf{u}$, will cause the process to drift to the origin along the directions orthogonal to $\text{Im}(\Gamma)$.

Under this parametrisation, the dynamics in the affine subspace simplify significantly,
\begin{align}
    \hat{\mathbf{f}}(\mathbf{y}) + \mathbf{y}& = W_s\mathbf{h}(\Gamma\mathbf{y}+\mathbf{b}) + I_s,
\end{align}
where we have moved the leak term to the left-hand-side for convenience. Here, we can notice the right-hand-side is equivalent to a two-layer perceptron with a $k$-neuron input layer, a $n-$neuron hidden layer, and a $k$-neuron output layer (Fig.~\ref{fig: framework}$b)$). The input-to-hidden weights and biases are given by $\Gamma$ and $\mathbf{b}$ respectively, whilst the hidden-to-output weights and biases are given by $W_s$ and $I_s$, and where $\mathbf{h}(\cdot)$ plays the role of the element-wise activation function. It is well-known that this parametrisation is a \textit{universal approximator} i.e. it can approximate any continuous function to arbitrary accuracy \cite{Hornik1989universalapprox}. Moreover, it can be efficiently optimised via backpropagation of error \cite{Goodfellow2016deeplearning}. 

The diffusion matrix in the affine subspace also simplifies to $B_sB_s^{\top}$. Here we assume correlated additive noise is injected into the RNN through the matrix $B$, which is carefully constructed to ensure the target isotropic diffusion over $\mathcal{A}$. {\color{black} This differs from the common assumption of isotropic noise at the level of neural activity, $B=\mathbf{I}_n$, and from the structured, non-isotropic noise correlations observed empirically, for example between similarly tuned neurons and aligned with stimulus-coding directions \cite{ebrahimi2022emergent}. Our $B$ does introduce correlations across neurons, but it is chosen so that the diffusion is isotropic over $\mathcal{A}$, which keeps the analysis in closed form rather than reflecting measured noise statistics. Such noise can be interpreted as shared input to the network (including stimulus variability), for which structured noise is more natural than for independent intrinsic fluctuations \cite{ebrahimi2022emergent}.}

{\color{black} From our extension to stochastic dynamics, we introduce the loss function,}
\begin{align}
\mathcal{L} &= ||\mathbf{f}(\mathbf{y})+ \mathbf{y} - W_s\mathbf{h}(\Gamma \mathbf{y}+ \mathbf{b}) - I_s|| + \lambda_{\text{diff}}||\sigma^2 \mathbf{I}_k -B_sB_s^{\top}||,
\end{align}
where $\lambda_{\text{diff}}$ is a hyper-parameter controlling the relative importance of the drift and diffusion losses, and where we optimise over the parameters $\{\Gamma, \mathbf{b}, W_s, I_s,B_s\}$, for a total of $2nk + n + k + kd$ free parameters.

{\color{black} Once the two-layer perceptron has been trained, the RNN can be reconstructed from the weights and biases to yield an autonomous neural circuit, in the form of an SDE, from which trajectories can be sampled using a stochastic integrator (App.~\ref{app: sim}). When trajectories are projected into the learnt affine subspace, they will follow the target SDE.}

\subsection{Symmetric-asymmetric decomposition of neural network dynamics}
\label{sec: symmetric-asymmetric decomp}

{\color{black} If we consider a low-rank but symmetric network, then it follows that the projected dynamics follow the generalised gradient of an energy potential,
\begin{align}
    \hat{\mathbf{f}}(\mathbf{y})&=-\mathbf{K}(\mathbf{y})\nabla_{\mathbf{y}}S(\mathbf{y}),
\end{align}
where $S(\mathbf{y}) = \tilde{E}(\Gamma \mathbf{y} + \mathbf{b})$ and $\mathbf{K}(\mathbf{y}) = \Gamma^{\dagger}\mathbf{D}(\Gamma \mathbf{y}+\mathbf{b})(\Gamma^{\dagger})^{\top}$ which is positive semi-definite.  $E$, $\tilde{E}$ and $\mathbf{D}$ are defined in App.~\ref{app: hopfield}, following Hopfield's seminal paper \cite{hopfield1984continuous}. This formulation motivates a decomposition of a trained RNN into symmetric and asymmetric components as any dynamics which go beyond monotonic decreases in energy must be encoded via the asymmetry of the connectivity matrix. }

We consider $W = W_{\text{sym}} + W_{\text{asym}}$, where the symmetric component, $W_{\text{sym}}$, descends an energy landscape, whilst the asymmetric component, $W_{\text{asym}}$, encodes rotational dynamics,
\begin{align}
\hat{\mathbf{f}}& = \hat{\mathbf{f}}_{\text{sym}}+ \hat{\mathbf{f}}_{\text{asym}},\\
\hat{\mathbf{f}}_{\text{sym}}(\mathbf{u}) &= -\mathbf{u} + W_{\text{sym}}\mathbf{h}(\mathbf{u}) + I,\notag\\
\hat{\mathbf{f}}_{\text{asym}}(\mathbf{u}) &= W_{\text{asym}}\mathbf{h}(\mathbf{u}).\notag
\end{align}
This decomposition has similarities to the Helmholtz-Hodge decomposition of a vector field \cite{GLOTZL2023127138} or SDE \cite{DaCosta_2023}, or the GENERIC decomposition of a stochastic process \cite{duong2023generic}. However, it is not identical to either, as the symmetric component is neither a conservative field, nor is it equal to the time-reversible component of the stochastic process. We consider decompositions based on time-irreversibility in more detail in Sec.~\ref{sec: attractor design} and \ref{sec: network hhd}.

At first glance, it is tempting to use the unique decomposition of the network into a symmetric and antisymmetric component,
\begin{align}
    W & = \frac{1}{2}\left(W+W^{\top}\right) + \frac{1}{2}\left(W-W^{\top}\right),\label{eq: decomp1}
\end{align}
where antisymmetric means $A=-A^{\top}$. However, with this decomposition, the affine subspace is no longer invariant under the individual drift components, $\hat{\mathbf{f}}_{\text{sym}}$ and $\hat{\mathbf{f}}_{\text{asym}}$. In order for the projected dynamics of the symmetric component to admit an energy function, $S(\mathbf{y})$, on $\mathbb{R}^k$, we require that $W_{\text{sym}}= \Gamma A$ for some $A$, which is not necessarily true despite $W=\Gamma W_s$. Using this decomposition, the projected dynamics of this symmetric component are not `attractor'-like and the projected process does not converge to the minimum of $S(\mathbf{y})$.

A more motivated decomposition can be achieved by splitting the network as,
\begin{align}
    W & = \Gamma(\Omega + \Pi),\label{eq: decomp2}
\end{align}
where $\Gamma\Omega = (\Gamma \Omega)^{\top}$ i.e. it is symmetric. Let us note that $\Gamma \Pi$ is asymmetric, unless $W=W^{\top}$, but not antisymmetric, except in the special case when this decomposition coincides with Eq.~(\ref{eq: decomp1}). Such a decomposition is clearly not unique, thus we define the `best' decomposition to be,
\begin{align}
    \Omega = \text{argmin}_{A \in \mathbb{R}^{k \times n}}||\Gamma A - C||_F^2,
\end{align}
where $C = \frac{1}{2}(W+W^{\top})$ i.e. our symmetric part is the most similar to the canonical decomposition, with respect to the Frobenius norm. This problem has a closed form solution (see App.~\ref{app: decomp}),
\begin{align}
    \Omega = \Gamma^{\dagger}C\Gamma \Gamma^{\dagger}.
\end{align}
This `improved' decomposition gives two drift components, where the affine subspace is invariant under each of them, and where $\hat{\mathbf{f}}_{\text{sym}}$ descends the energy function $S(\mathbf{y})$ associated with the network $W_{\text{sym}}= \Gamma \Omega$.

\subsection{Pipeline and examples}
\label{sec: examples 1}

{\color{black}To illustrate the training framework and the network decomposition, we first consider three well-known nonlinear systems with isotropic additive noise: the stochastic \textit{van der Pol} (VDP) system, \textit{Lorenz} system (LS), and \textit{Dadras} system (DS) (defined in App.~\ref{app: nonlinear examples})}. We train an RNN with 64, 512, and 1024 neurons for each of the systems respectively with backpropagation. Using the fitted parameters, we integrate the RNN as an autonomous SDE using the Euler-Maruyama scheme (see App.~\ref{app: sim}). Projecting the dynamics into the learnt affine subspace, we recover trajectories from the target SDE. Next, we decompose the RNN connectivity as described in Sec.~\ref{sec: symmetric-asymmetric decomp}, where we can investigate which features of the dynamics are encoded in each part of the network respectively.
\begin{figure*}
    \centering
    \includegraphics[width=0.95\linewidth]{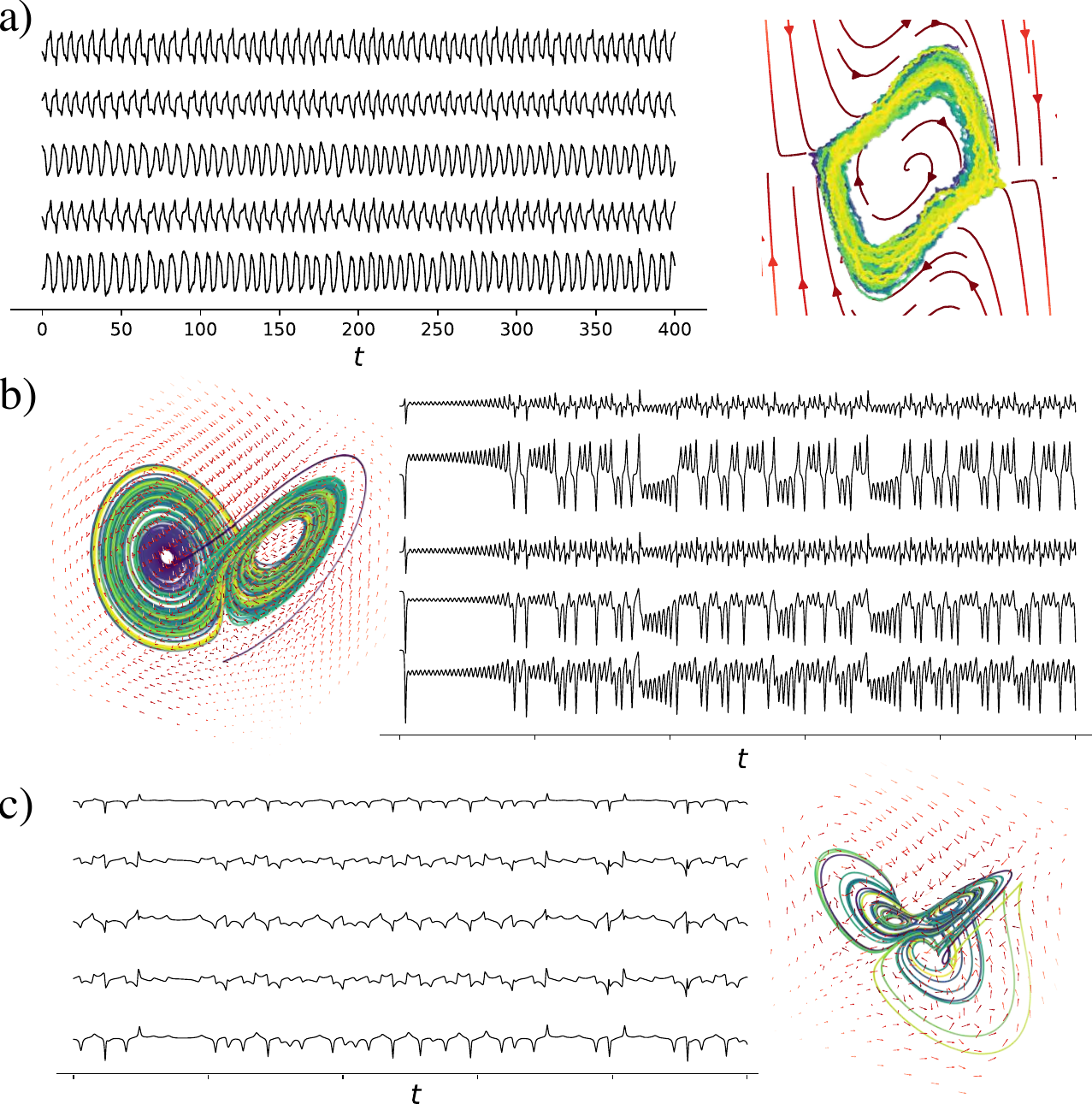}
    \caption{\textbf{Embedding attractors in neural networks.} $a)$ Traces from 5 randomly selected neurons in a 64-neuron RNN trained to embed a stochastic VDP oscillator in a latent subspace. In the subspace, we see the characteristic VDP limit-cycle. $b-c)$ Traces from a 512-neuron RNN trained to embed the LS, and a 1024-neuron RNN trained to embed the DS. In panel $a)$ the arrows represent streamlines of the vector field, whilst in $b-c)$ they represent the vector field itself. The colour of the arrows represents the magnitude of the drift at that point.}
    \label{fig: embedding attractors}
\end{figure*}

Fig.~\ref{fig: embedding attractors} shows example traces from 5 randomly selected neurons in each of the RNNs, alongside the phase-plane dynamics in the learnt latent space. We see that the system is able to encode all three nonlinear systems, and that the trajectories of individual neurons mirror the qualities of the embedded attractor. 

\begin{figure*}
    \centering
    \includegraphics[width=0.9\linewidth]{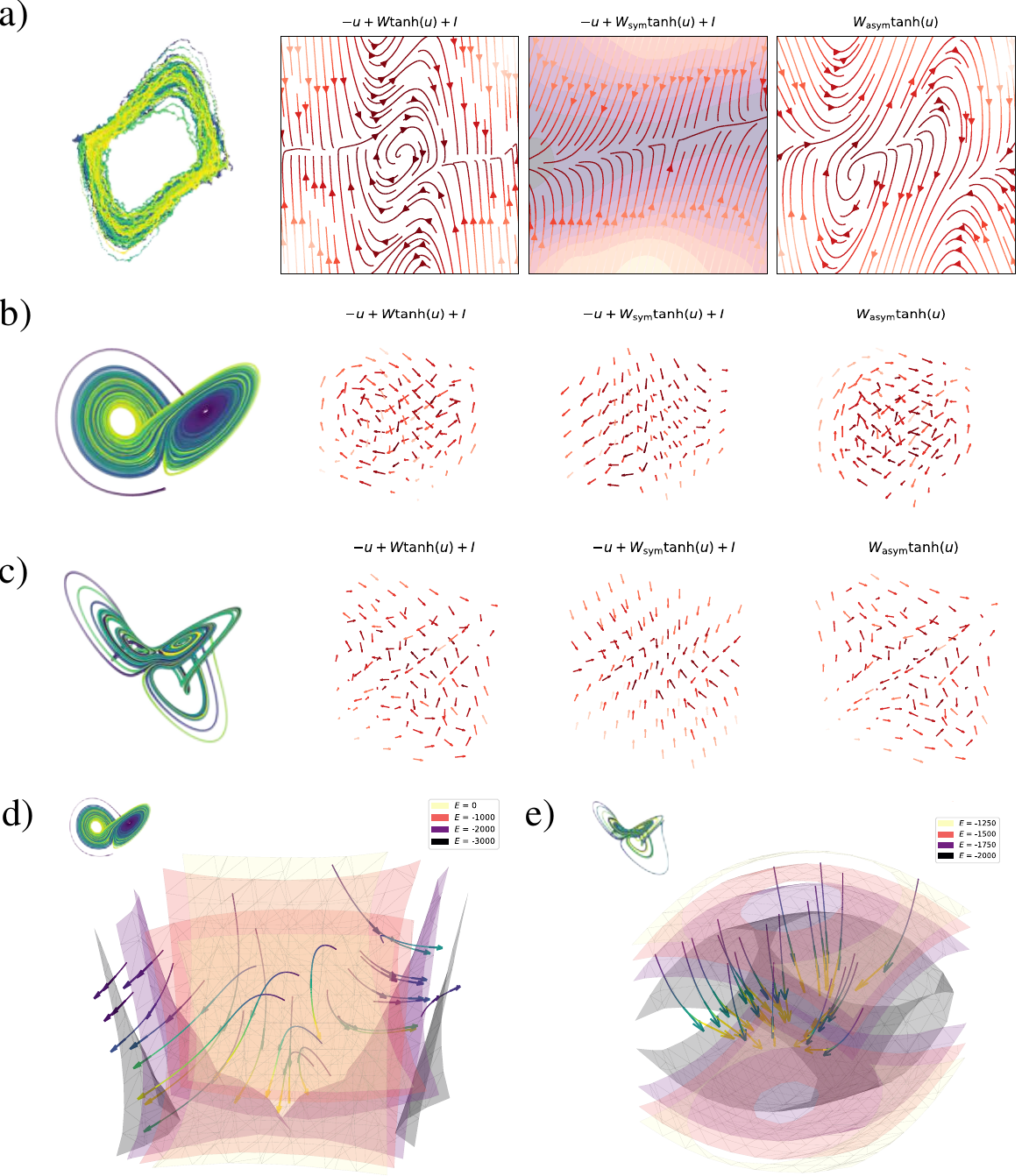}
    \caption{\textbf{Symmetric-asymmetric decomposition of RNNs.} $a)$ Decomposition of RNN trained to encode the VDP. The symmetric dynamics descend the contoured energy function. This appears to keep the process near a centreline, whilst the asymmetric component provides rotational dynamics. $b-c)$ Decompositions of RNNs trained to encode the LS and DS. For the DS, the symmetric dynamics keep the process near the attractor, whilst the asymmetric component provides rotational dynamics. For the LS,  it appears the symmetric component pushes the dynamics away from the origin, and the asymmetric component provides a restoring force. $d-e)$ This is confirmed by plotting isopotentials of the energy functions and their associated streamlines for the LS and DS respectively.}
    \label{fig: chaos decomp}
\end{figure*}

{\color{black} Fig.~\ref{fig: chaos decomp} shows the results of the decomposition for the VDP, LS,  and DS examples. This decomposition gives us two vector field components, one of which is the generalised gradient of an associated energy function. Panels $a-c)$ show streamlines of both components for each system, whilst Panels $d-e)$ show characteristic isopotentials of the energy function for the LS and DS systems with streamlines. Studying the energy landscapes, we make the following observations: the symmetric component of the VDP system pushes the process near the centreline of the attractor, then diverging towards infinity; the symmetric component of the DS pushes the system towards a single fixed point (an energy minimum); the symmetric component of the LS drives the process away from the attractor, diverging towards the corners of phase space. For each system, the asymmetric, rotational component provides a counter-balancing force that prevents the system from either diverging (VDP, LS) or arriving at fixed point (DS).}

\section{Switching and cycling between attractors}
\label{sec: attractor design}

Attractor dynamics are frequently cited as both emergent features in neural data, and as a possible mechanism for computation via dynamics \cite{Amit1989modelingbrain, Khona2022attractor}. These attractors are most easily understood in terms of energy landscapes, where the dynamics perform some kind of gradient descent, arriving at a local minimum in the form of a fixed point or other low-dimensional manifold, e.g. a line or a ring. On the one hand, a range of computational mechanisms, both neuroscientific and other, rely on such dynamics, including stochastic gradient descent \cite{Goodfellow2016deeplearning}, the `free-energy principle' \cite{Friston2011valueattractors}, and the class of so-called energy-based models \cite{LeCun2007energybased}. On the other hand, in both neural and biological systems, asymmetry and non-reciprocity are ubiquitous, typically disrupting the dynamics associated with symmetric models and their energy landscapes \cite{yan2013landscape,nartallokalu2025review,nartallokaluarachchi2024broken,fruchart2021nonreciprocal, Ninou2025Curl}. However, designing learning systems that utilise such dynamics has, so far, been evasive.

Based on our method to train an RNN to encode a target SDE in a latent affine subspace, we now introduce two classes of SDEs that are defined by energy potential gradients and thus have attracting dynamics. The first of these uses network inputs to \textit{switch} between the minima of the energy potential. Here the network inputs represent sensory information or the output of a disjoint neural circuit, which prompts the RNN to retrieve a particular memory in light of this information {\color{black} i.e. associative memory.} The second is a class of stationary diffusions that {\color{black} are crafted to} use irreversible, rotational currents to cycle between minima in a preferred direction. In this case, our model attempts to describe the ability of a memory system to autonomously transition between a series of a related memories in a preferred direction. We show that both these dynamics can be encoded within an RNN, as illustrative examples of how a neural circuit could encode more complex forms of associative memory.

Before progressing, we introduce some preliminaries for stochastic dynamics and their nonequilibrium steady-states. The general form for SDEs which we will consider is given by,
\begin{align}
    d\mathbf{y}(t) = \mathbf{f}(\mathbf{y})\;dt + {\color{black} K} \;d\mathbf{w}(t),
\end{align}
with $\mathbf{y}(t)\in \mathbb{R}^k$ and $K \in \mathbb{R}^{k \times d}$. The diffusion matrix is defined by ${\color{black} D = KK^{\top}/2}$ which is positive semidefinite.\footnote{{\color{black} This general form for an SDE with additive noise includes both the target processes, where we restrict to isotropic noise, and the dynamics of the RNN itself, where the noise is correlated.}} We will focus on cases where the drift $\mathbf{f}$ is designed to encode descent of an energy landscape and, in the second case, an additional rotational flow between minima. The \textit{Fokker-Planck} (FP) equation describes the dynamics of the probability density, $p(\mathbf{y},t)$,
\begin{align}
    \partial_tp & = -\nabla \cdot \mathbf{J},\\
    \mathbf{J}(\mathbf{y},t)& = \mathbf{f}(\mathbf{y})p(\mathbf{y},t) - D\nabla p(\mathbf{y},t),\notag
\end{align}
where $\mathbf{J}$ is the probability flux \cite{pavliotis2014stochproc}. The SDE is ergodic with stationary density $\pi(\mathbf{y})$ if $p(\mathbf{y},t)\rightarrow_{t\rightarrow \infty}\pi(\mathbf{y})$ and $\nabla \cdot \mathbf{J}_{\text{ss}} = 0$, where $\mathbf{J}_{\text{ss}}$ is the flux with respect to the density $\pi$.

A stationary process is said to be an \textit{equilibrium steady-state} (ESS) if $\mathbf{J}_{\text{ss}}=0$, which is also known as the \textit{detailed balance condition}, and implies that the process is time-reversible \cite{Jiang2004noneq}. Otherwise, the process is in a \textit{nonequilibrium steady-state} (NESS), is time-irreversible, and produces entropy. We also introduce the \textit{Helmholtz-Hodge decomposition} (HHD) of the process, which decomposes the drift of a stationary diffusion into two components,
\begin{align}
\label{eq: hhd}
    \mathbf{f}& = \mathbf{f}_{\text{rev}} + \mathbf{f}_{\text{irr}},\hspace{10pt}
    \mathbf{f}_{\text{rev}}= D\nabla \log \pi,\hspace{10pt}
    \mathbf{f}_{\text{irr}} = \mathbf{J}_{\text{ss}}/\pi,
\end{align}
where $\mathbf{f}_{\text{rev}}$ is a time-reversible (shifted) gradient flow and $\mathbf{f}_{\text{irr}}$ is the time-irreversible circulation \cite{DaCosta_2023}. Finally, the degree of irreversibility can be quantified by the \textit{entropy production rate} (EPR), given by,
\begin{align}
\label{eq: EPR}
    \Phi & = \int_{\mathbb{R}^k}(\mathbf{f}_{\text{irr}})^{\top}D^{-1}\mathbf{f}_{\text{irr}}\pi \;d\mathbf{y},
\end{align}
which is non-negative, and zero if and only if the process is in detailed balance \cite{Jiang2004noneq}.\footnote{When $D$ is not full-rank, we can use its MP inverse, provided certain conditions are met (see Sec.~\ref{sec: network hhd} and Ref.~\cite{DaCosta_2023}).} 

\subsection{Input-driven attractor-switching}
\label{sec: switching}

We consider an RNN of the form,
\begin{align}
    d\mathbf{u}(t) & = \mathbf{F}(\mathbf{u}) + G\mathbf{s}(t)\;dt + B \;d\mathbf{w}(t),\\
    F_i & = -u_i + \sum_{j}W_{ij}v_j + I_i + \Gamma \mathbf{d}, \notag
\end{align}
where we have an additional \textit{input} term $G\mathbf{s}(t) + \Gamma\mathbf{d}$, where $\mathbf{s}(t) \in \mathbb{R}^l$ is the low-dimensional input signal, and $G\in \mathbb{R}^{n \times l}$ represents its projection into the RNN. Moreover, we assume that both the input connectivity, $G = \Gamma G_s$, and the input bias, $\Gamma \mathbf{d}$, are low-rank, thus the dynamics remain constrained to the affine subspace.

\begin{figure}
    \centering
    \includegraphics[width=\linewidth]{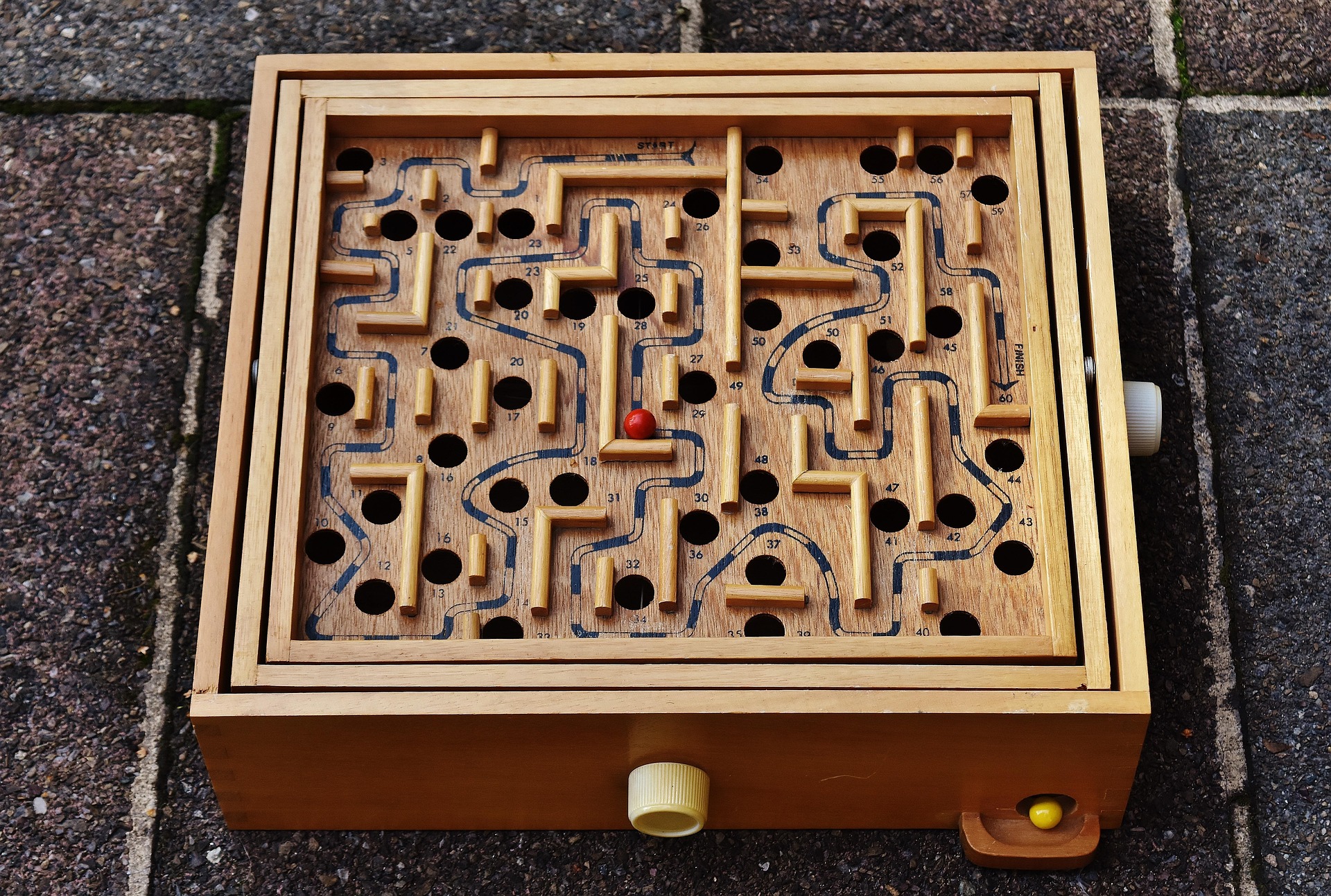}
    \caption{\textbf{`Tilting' the energy landscape.} Our approach uses network inputs to `tilt' the energy landscape, encouraging the process to push a trajectory into a particular energy minimum, by making it a global minimum. This is a soft version of the child's `labyrinth' game, where the board can be tilted such that the ball rolls into a specific hole -- though in the actual game, one tries rather hard to avoid them.}
    \label{fig: labyrinth}
\end{figure}

Next, we assume that our network, in the absence of inputs, encodes a gradient flow on a energy potential, $V:\mathbb{R}^k\rightarrow\mathbb{R}$, with a number of minima of approximately equal energy, and isotropic noise,
\begin{align}
    d\mathbf{y}(t) = -\nabla V(\mathbf{y})\;dt + \sigma \;d\mathbf{w}(t).
\end{align}
Typically trajectories will be attracted to the closest minimum, eventually exploring the landscape via random diffusion. It is well-known that this process has a stationary \textit{Boltzmann density},
\begin{align}
    \pi(\mathbf{y}) & = \frac{1}{Z}\exp\left(-\frac{2V(\mathbf{y})}{\sigma^2}\right),
\end{align}
and is in an ESS \cite{pavliotis2014stochproc}. Our aim is to direct the system towards a particular minimum using the network input. The dynamics on the affine subspace can be written as,
\begin{align}
    d\mathbf{y}(t) & = -\nabla V(\mathbf{y}) + G_s\mathbf{s}(t) + \mathbf{d} \; dt + \sigma \;d\mathbf{w}(t).
\end{align}
Considering a constant input over time, and defining $\mathbf{c} = G_s\mathbf{s} + \mathbf{d}$, we notice that the drift can be written as the gradient of a modified potential function, $\mathbf{f}(\mathbf{y}) = - \nabla[V(\mathbf{y})-\mathbf{c}\cdot \mathbf{y}]$. Adding a term that is constant in $\mathbf{y}$ corresponds to `tilting' the potential to prioritise a particular minimum, making it a global minimum. This approach can be thought of in analogy with a child's `labyrinth' game (Fig.~\ref{fig: labyrinth}), where the objective is to tilt the board, which serves as the energy potential, such that the ball rolls into a particular hole i.e. the chosen energy minimum.

 Assuming that the potential has a set of $m$ deep, well-separated minima, $\{\bm{\mu}_j\}_{j=1}^m$, for each minima there is a convex subset $\mathcal{C}_{j} \subset \mathbb{R}^k$, where $\mathbf{c} \in \mathcal{C}_{j}$ makes $\bm{\mu}_j$ the global minimum. More explicitly, defining $V^0_j = V(\bm{\mu_j})$, if $\mathbf{c}\cdot(\bm{\mu}_i^*-\bm{\mu}_j)>V^0_{i^*}-V^0_j$ for all $j\neq i^*$, then $\bm{\mu}_i^*$ is the global minimum, assuming the perturbation is small enough. Given values of $\mathbf{c}$ for each minimum, it remains to compute $(G_s, \mathbf{d})$ for a particular choice of coding scheme for the inputs. Example schemes include `one-hot' or random Gaussian inputs. In App.~\ref{app: computing c}, we provide a concrete method for picking values of $\mathbf{c}$ with quadratic programming, followed by computing an appropriate $G_s$ and $\mathbf{d}$ with ridge regression.
\begin{figure*}
    \centering
    \includegraphics[width=\linewidth]{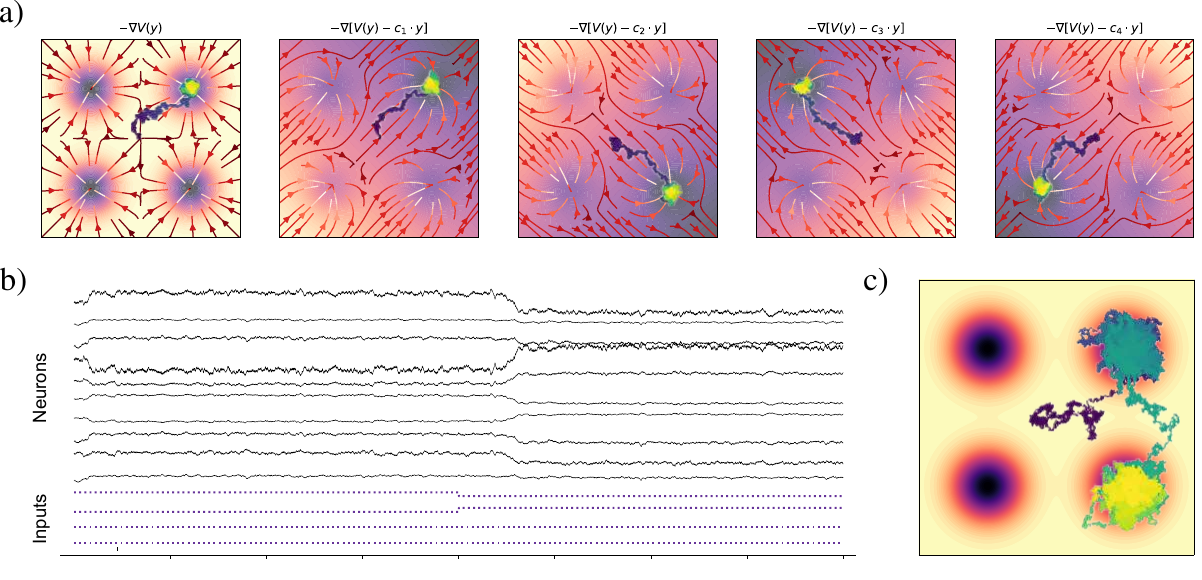}
    \caption{\textbf{Input-driven attractor-switching}. $a)$ We consider a four-well energy potential in the plane. In the absence of inputs, the trajectories converge to a minima at random, with a preference for the closest one. We can then tilt the potential with an input, which prioritises the relevant minima, and causes the trajectories to tend towards it. $b)$ We train a 256-neuron RNN to embed the potential-gradient dynamics. Using a one-hot coding for the input, we learn the coupling matrix input connectivity and bias, then add the input to the RNN during integration. We start with the input that drives the process to the top-right well, then switch it halfway through to the bottom-right well. We plot traces from 10 random neurons and see the network dynamics react to the input and converge to the other minima. $c)$ Projecting the RNN trajectory onto the latent space, we see the attractor-switching that occurs due to the change in input.}
    \label{fig: switching}
\end{figure*}

We now consider a numerical example. As the energy potential, we choose a sum of Gaussian wells, 
\begin{align}
    V(\mathbf{y}) & =- \sum_{j=1}^ma_j\exp\left(-\frac{||\mathbf{y}-\bm\mu_j||^2}{2\nu_j^2}\right),
\end{align}
where the $m$ minima can be placed at arbitrary positions $\bm{\mu}_j \in \mathbb{R}^k$, and we can specify the depth, $a_j$, and breadth, $\nu_j$, of each well. As shown in Fig.~\ref{fig: switching}, we place four identical minima at points in the plane. In the absence of inputs, the process typically converges to the minima that is closest. However, activating the corresponding input code, causes the potential to tilt and encourages trajectories to converge to a specified minimum. We train a 256-neuron RNN to encode the potential-gradient, then, using a one-hot coding for the inputs, we fit the input-connectivity and bias, $G$ and $\mathbf{d}$ (see App.~\ref{app: computing c} for details). We then sample a trajectory from the RNN, where the input signal switches at the halfway point from one minimum to another. Panels $b)$ and $c)$ of Fig.~\ref{fig: switching} show traces of 10 random neurons in the network, and the projection of the network dynamics into the latent space, respectively. We see that the process converges to the first attractor, but then switches to the second due to the change in the input. {\color{black} Sussillo and Barak consider a related approach for transitioning between fixed points of a deterministic RNN using input signals \cite{Sussillo2013opening}.}

\subsection{Attractor-cycling stationary diffusions}
\label{sec: cycling}

In this section, we design a class of planar SDEs, which not only converge to energy minima, but also cycle between these minima in a preferred direction using irreversible currents {\color{black} in an NESS}. As before, we begin with an energy potential and define a gradient flow,
\begin{align}
    d\mathbf{y}(t) = -\nabla V(\mathbf{y})\;dt + \sigma \;d\mathbf{w}(t).
\end{align}
{\color{black} The energy landscape ensures the attractor-like dynamics of associative memory. Next, we add in a rotational component, resulting in a process of the form,}
\begin{align}
    d\mathbf{y}(t) = -\nabla V(\mathbf{y}) + \mathbf{R}(\mathbf{y})\;dt + \sigma \;d\mathbf{w}(t).    
\end{align}
{\color{black} To ensure that the rotational dynamics do not disrupt the potential gradient, and thus the attracting states, we enforce that the process with rotational dynamics must still converge to the Boltzmann density.} It is easy to show that the condition for this to be satisfied is that $\nabla \cdot (\mathbf{R}\pi) = 0$ i.e. the rotational component is \textit{divergence-free with respect to the stationary density} (DFSD). This condition is closely related to the HHD defined previously. In this case, we have simply that $\mathbf{f}_{\text{rev}} = -\nabla V$ whilst $\mathbf{f}_{\text{irr}} = \mathbf{R}$. By adding on a rotational component that does not disrupt the Boltzmann density, we construct stationary diffusions that descend the energy landscape whilst being driven rotationally by stationary probability flux. In App.~\ref{app: high dim cycling}, we construct both high-dimensional and non-coplanar examples. Specifically we design a process that can cycle between a pair of pixelated images in $\mathbb{R}^{1024}$, and another that cycles between minima placed on a saddle in $\mathbb{R}^3$, although we do not encode these in an RNN. {\color{black} We now focus on crafting planar examples by designing $\mathbf{R}$ with the desired properties, which we then embed within an RNN.}

\begin{figure}
    \centering
    \includegraphics[width=\linewidth]{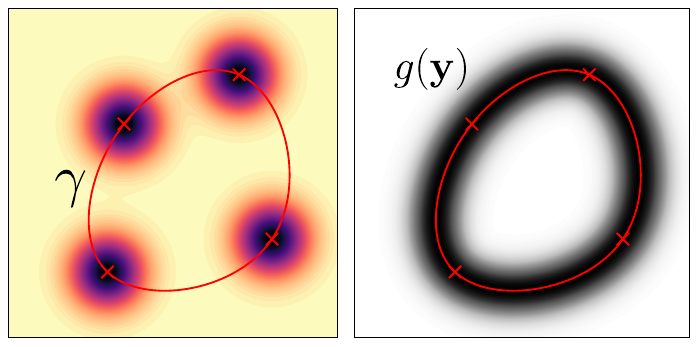}
    \caption{\textbf{Constructing rotational diffusions.} To construct an SDE with rotation between energy minima, we begin with the gradient of the energy potential. We then draw a closed curve, $\gamma$, through the minima in the direction of rotation. From this, we can define a rotational drift which decays with distance from the curve due to the weighting kernel $g(\mathbf{y})$.}
    \label{fig: curve}
\end{figure}
In $\mathbb{R}^2$, or in the case that the attractors are coplanar in plane $P$, we can construct a differentiable, closed planar curve, $\gamma$, which connects points in the order that we would like the field to rotate (Fig.~\ref{fig: curve}). Next, we design a rotational field that provides a force in the direction of this closed curve, acting along all concentric curves with the same shape. For example, if the curve is a circle, this is the set of circles with the same centre but different radii. To do this, we consider $\gamma = \partial O$ to be the boundary of a bounded planar region $O \subset P$, and define the \textit{signed distance} to be,
\begin{align}
    \phi(\mathbf{y}) & =  \left\{\begin{matrix}
-\text{dist}(\mathbf{y},\gamma) & \text{if }  \mathbf{x}\in O \\
\text{dist}(\mathbf{y},\gamma) & \text{if }  \mathbf{x}\in P \backslash O
\end{matrix}\right.
\end{align}
where $\text{dist}(\mathbf{y},\gamma) = \inf_{\mathbf{p}\in \partial O} ||\mathbf{y}-\mathbf{p}||$. This defines a potential where each level set is a closed curve with the same shape, up to an inflation or deflation, as $\gamma$. We would like our rotational field to be tangent to these level sets. We obtain such a field by rotating $\nabla \phi$ by $\pi/2$ via an antisymmetric matrix $\mathbf{M}$, which in $\mathbb{R}^2$ is given by,
\begin{align}
    \mathbf{M} = \begin{pmatrix} 0 & 1\\
    -1 & 0
    \end{pmatrix}.
\end{align}
{\color{black} By itself, this rotational field would dominate the gradient dynamics far from the minima, which we avoid by introducing a weighting kernel,}
\begin{align}
\label{eq: weighting kernel}
    g(\mathbf{y}) & = \exp \left(-\frac{\phi^2(\mathbf{y},\gamma)}{2\xi^2}\right),
\end{align}
which decays exponentially with the distance from the closed curve $\gamma$ at a rate inversely proportional to $\xi^2$ (Fig.~\ref{fig: curve}). {\color{black} Putting together the components, we have a vector field,
\begin{align}
    \bm{\zeta}(\mathbf{y}) & = g(\mathbf{y})\mathbf{M}\nabla\phi(\mathbf{y}),
\end{align}
which is divergence-free (see App.~\ref{app: coplanar DFSD}). Finally, we divide through by $\mathbf{\pi}$ and introduce the speed of rotation $\alpha$ to obtain the rotational vector field,
\begin{align}
    \mathbf{R}(\mathbf{y}) & = \frac{\alpha}{\pi(\mathbf{y)}}g(\mathbf{y})\mathbf{M}\nabla\phi(\mathbf{y}),
\end{align}
where $\mathbf{R}$ is DFSD (see App.~\ref{app: coplanar DFSD}). This resulting drift is given by,} 
\begin{align}
    \mathbf{f} = -\nabla V + \mathbf{R},
\end{align}
that has both attractive and cycling dynamics. Given a set of points $\{\bm{\mu}_j\}$ to encode as minima of a Gaussian potential, the parameters $\{\{a_j\},\{\nu_j\}, \sigma, \xi, \alpha\}$, can all be chosen to obtain the desired behaviour in the process (details on parameters are given in App.~\ref{app: params}).

\begin{figure*}[ht]
    \centering
    \includegraphics[width=0.95\linewidth]{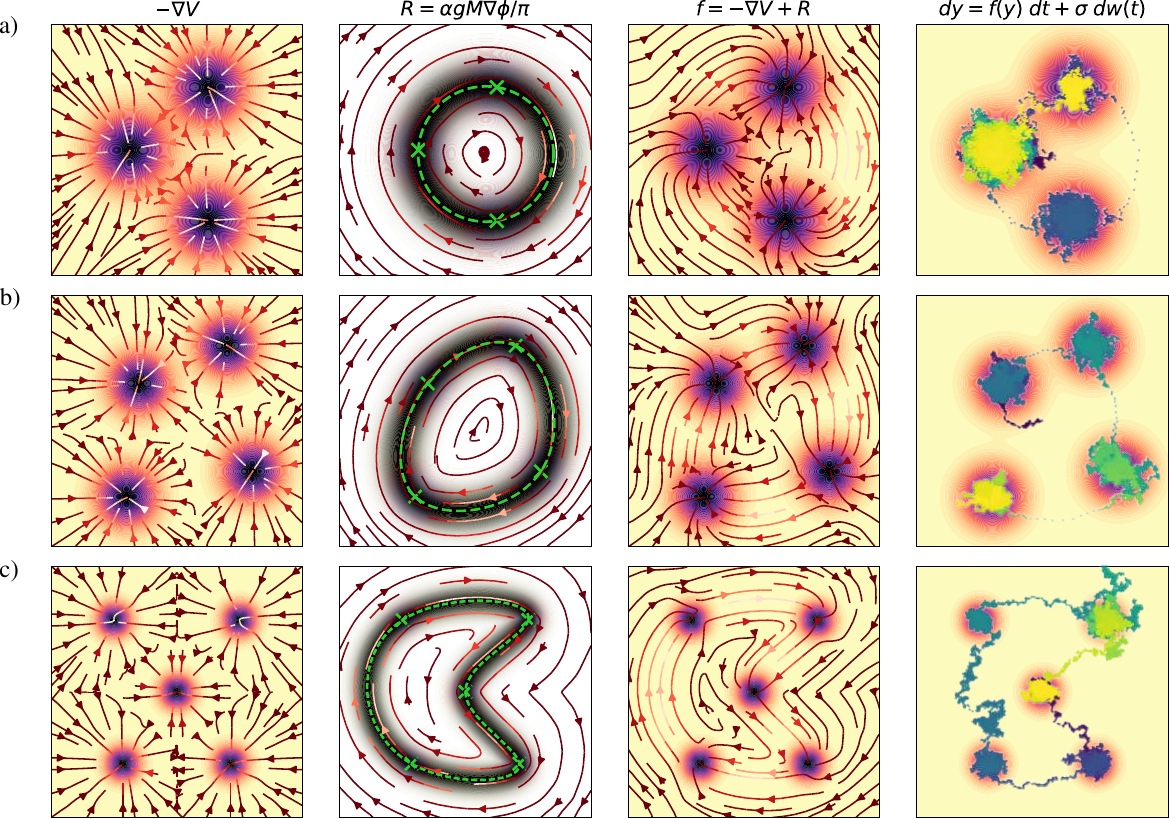}
    \caption{\textbf{Attractor-cycling diffusions.} We construct example diffusion processes that cycle 3, 4, and 5 attractors with irreversible currents (Panels $a-c$). These are composed of a gradient flow (first column) plus a rotational component. The rotational component is constructed by taking a smooth curve through the minima, defining a potential as the distance from the curl, multiplying by a decaying weight kernel (contour in second column), and dividing by the stationary density (second column). Combined, this defines a vector field that both converges to, and cycles between attractors (third and fourth column).}
    \label{fig: cycling}
\end{figure*}
\begin{figure*}[ht]
    \centering
    \includegraphics[width=0.7\linewidth]{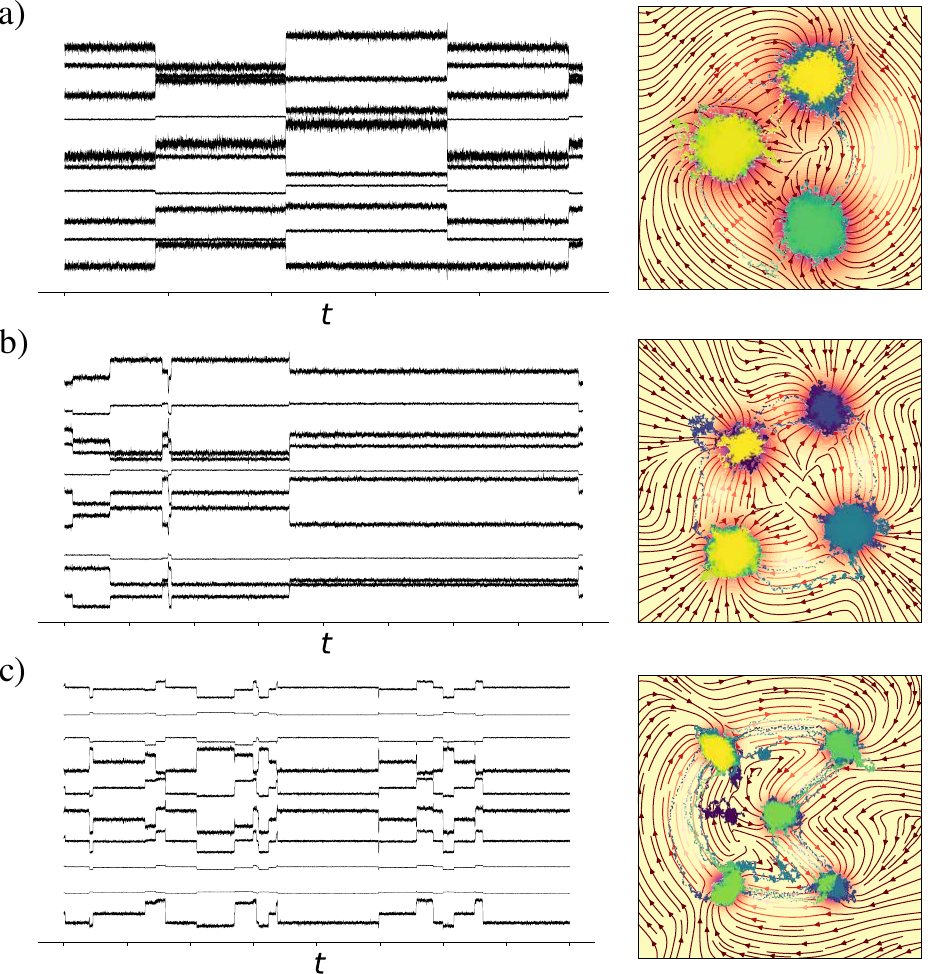}
    \caption{\textbf{Attractor-cycling in RNNs.} We train 128-neuron RNNs to approximate the planar diffusion processes from Fig.~\ref{fig: cycling} for $a)$ three, $b)$ four, and $c)$ five minima respectively. Trajectories from the RNN can then be sampled autonomously, and the learnt drift can be projected onto the learnt affine subspace. We find that the RNNs are able to learn the target diffusion processes. For each RNN, we plot traces from ten random neurons. We see that the attractor cycling dynamics in the latent subspace are visible in the raw trajectories.}
    \label{fig: rnn_cycling}
\end{figure*}

We construct three examples of diffusion processes with three, four, and five Gaussian wells placed in the plane. In the case of the three minima we choose the curve $\gamma$ to be the intersecting circle, whilst for four and five minima we compute it numerically with cubic splines (see App.~\ref{app: splines}). Fig.~\ref{fig: cycling} illustrates the example processes for $a)$ three, $b)$ four, and $c)$ five minima. Next, we train a different 128-neuron RNN to approximate each diffusion process in a latent affine subspace. Once the model is trained, we sample trajectories from the RNN and project them into the learnt subspace. Fig.~\ref{fig: rnn_cycling} shows example traces from ten randomly selected neurons in each network, as well as a projection of the trajectory and the drift into the affine subspace. We see that the RNN is able to encode such autonomous attractor-cycling dynamics in a latent subspace. Moreover, as in the case of the nonlinear systems (Fig.~\ref{fig: embedding attractors}), the activity of individual neurons hints at the character of the encoded process i.e. here we see neuronal activity jumping between piecewise approximately constant segments. 

\section{Nonequilibrium steady-states of neural networks}
\label{sec: network hhd}

Computing properties of the NESS for a high-dimensional network is particularly challenging, except in the case of simple, solvable processes \cite{nartallokaluarachchi2024broken}, as the FP equation is high-dimensional and its numerical solution is computationally intractable. However, in the case of a high-dimensional RNN encoding a low-dimensional stationary diffusion in an invariant subspace, the situation simplifies significantly.

We consider, as before, an RNN with state $\mathbf{u}(t) \in \mathcal{A}$, where the latent dynamics follow the SDE,
\begin{align}
    d\mathbf{y}(t) = \mathbf{f}(\mathbf{y})\;dt + \sigma\;d\mathbf{w}(t),
\end{align}
which has a stationary density $\pi_{\mathbf{y}}(\mathbf{y})$. The stationary density of $\mathbf{u}$ is given by the \textit{pushforward measure} \cite{Bogachev2007measure},\footnote{From this point on, we are implicitly constraining $\mathbf{u}\in \mathcal{A}$ to avoid Dirac deltas in the expression.}
\begin{align}
    \pi_{\mathbf{u}}(\mathbf{u}) & = \frac{\pi_\mathbf{y}(\Gamma^{\dagger}(\mathbf{u}-\mathbf{b}))}{\sqrt{\det(\Gamma^{\top}\Gamma)}}.
\end{align}
Given the stationary density we can compute the HHD, stationary flux, and EPR -- it is closely related between the latent and network-level. Assuming that the HHD of the latent process is given by $\mathbf{f} = \mathbf{f}_{\text{rev}} + \mathbf{f}_{\text{irr}}$, we can decompose the network drift, $\mathbf{F}$ as,
\begin{align}
\label{eq: HHD network dynamics}
\mathbf{F}& = \mathbf{F}_{\text{rev}} + \mathbf{F}_{\text{irr}},\\
    \mathbf{F}_{\text{irr}}(\mathbf{u}) & = \Gamma \mathbf{f}_{\text{irr}}(\Gamma^{\dagger}(\mathbf{u}-\mathbf{b})),\notag\\
    \mathbf{F}_{\text{rev}}(\mathbf{u}) & = \Gamma \mathbf{f}_{\text{rev}}(\Gamma^{\dagger}(\mathbf{u}-\mathbf{b}))\notag.
\end{align}
{\color{black}
Similarly, the flux is related by,
\begin{align}
    \mathbf{J}_{\mathbf{u}}(\mathbf{u}) & = \frac{\Gamma \mathbf{J}_{\mathbf{y}}(\Gamma^{\dagger}(\mathbf{u}-\mathbf{b}))}{\sqrt{\det\left(\Gamma^{\top}\Gamma\right)}}.
\end{align}
Finally, in order to compute the EPR, we note that the traditional formula for EPR, Eq.~(\ref{eq: EPR}), assumes that the diffusion matrix is full-rank \cite{Jiang2004noneq}, which does not hold in our case where $2D = BB^{\top}=\Gamma B_s B_s^{\top}\Gamma^{\top}$, which is a symmetric positive semidefinite matrix with rank $k$. In this case, the EPR formula is given by,
\begin{align}
    \Phi_{\mathbf{u}} & = \int_{\mathbb{R}^n}\mathbf{F}^{\top}_{\text{irr}}D^{\dagger}\mathbf{F}_{\text{irr}}\pi_{\mathbf{u}}\;d\mathbf{u},
\end{align}
i.e. where $D^{-1}$ has been replaced by the MP inverse, if and only if $\mathbf{F}_{\text{irr}}(\mathbf{u}) \in \text{Range}(B(\mathbf{u}))$ where $B=\Gamma B_s$ \cite{DaCosta_2023}. Eq.~(\ref{eq: HHD network dynamics}) guarantees that $\mathbf{F}_{\text{irr}}$ lies in the image of $\Gamma$, and as we consider additive noise, it is sufficient for $B_s$ to be of full row-rank $k$. A necessary requirement for this is that $d \geq k$ i.e. we have at least as many independent noise-sources in the network as dimensions in the latent process. When $B_s$ has full row-rank, we have that the EPR of the latent process and the full network are identical,
\begin{align}
    \Phi_{\mathbf{u}} = \Phi_{\mathbf{y}}.
\end{align}
}
\begin{figure*}
    \centering
    \includegraphics[width=0.85\linewidth]{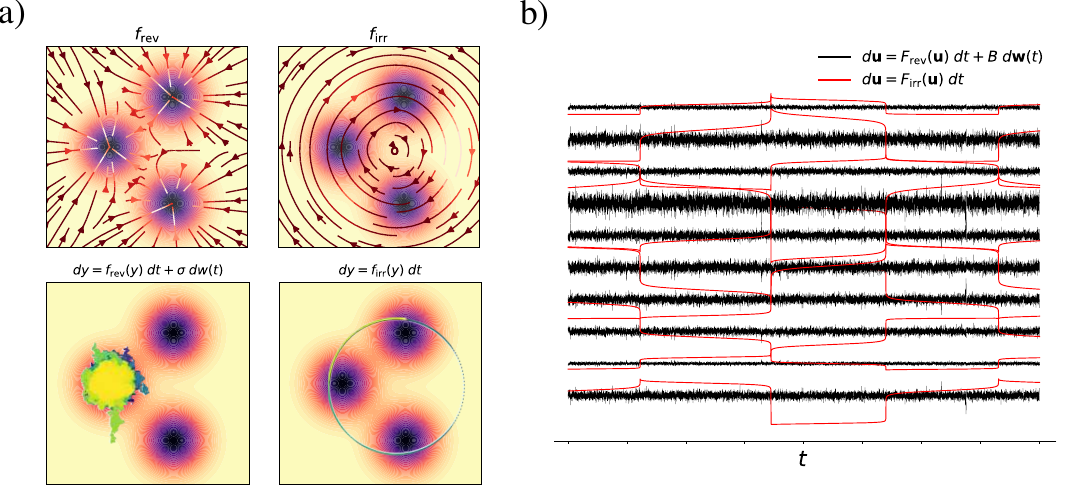}
    \caption{\textbf{The HHD of a latent, embedded process.} $a)$ When we embed a process with a known NESS, we can perform the HHD. In this example with three minima, we see that the reversible dynamics balance the diffusion and maintain the process on the stationary distribution, whilst the rotational dynamics create rotation around the curve $\gamma$. $b)$ We train an RNN with 64 neurons to approximate the process in a latent affine subspace. Using the HHD and Eq.~(\ref{eq: HHD network dynamics}), we can define the irreversible, $\mathbf{F}_{\text{irr}}$ and reversible, $\mathbf{F}_{\text{rev}}$ components of the high-dimensional dynamics. We show the reversible and irreversible activity of 10 random neurons over time. We see that the reversible dynamics at the network level keep the process near the minima, which results in trajectories that fluctuate noisily around a fixed point (an energy minimum), whilst the irreversible dynamics show limit-cycle behaviour.}
    \label{fig: hhd 1}
\end{figure*}

We now return to the example process with three minima given in Sec.~\ref{sec: cycling}. As the process has a solvable NESS, we can compute the HHD, $\mathbf{f}_{\text{rev}} = -\nabla V$ and $\mathbf{f}_{\text{irr}} = \mathbf{R}$, as shown in Panel $a)$ of Fig.~\ref{fig: hhd 1}. The reversible component balances the diffusion to maintain the process at stationarity, as seen in the example trajectory, whilst the irreversible component cycles around the curve $\gamma$. We then train a 64-neuron RNN to approximate the process in a latent affine subspace. Using the HHD of the example and Eq.~(\ref{eq: HHD network dynamics}), we define a decomposition of the high-dimensional dynamics in the ambient space, and sample trajectories from each component. Panel $b)$ shows a trajectory from each of these components. As with the dynamics at the latent level, we see that the reversible dynamics keep the process near the minima, balancing the noise, whilst the irreversible dynamics show limit-cycle behaviour.

\subsection{A Helmholtz-Hodge decomposition of the network}
Whilst Eq.~(\ref{eq: HHD network dynamics}) is a decomposition of the high-dimensional dynamics, it is challenging to interpret, as each component is a high-dimensional vector field that we cannot visualise. Moreover, the decomposition is not related directly to the learnt network connectivity or input currents, only to the learnt affine subspace.

Instead, we now relate these components to a decomposition of the network itself. We assume that the HHD of the network can be written as,
\begin{align}
\label{eq: RNN HHD 1}
    \mathbf{F}_{\text{irr}}(\mathbf{u}) & = -\varrho\mathbf{u }+W_{\text{irr}}h(\Gamma \mathbf{y} + \mathbf{b}) + I_{\text{irr}},\\
    \mathbf{F}_{\text{rev}}(\mathbf{u}) & = -(1-\varrho)\mathbf{u} + W_{\text{rev}}h(\Gamma \mathbf{y} + \mathbf{b}) + I_{\text{rev}},\notag
\end{align}
where $W = W_{\text{rev}} + W_{\text{irr}}$, $I = I_{\text{rev}} + I_{\text{irr}}$, and $\varrho\in [0,1]$. We also assume that these parts respect the parametrisation,
\begin{align}
W_{\text{rev}} &= \Gamma W_{s,\text{rev}}, && W_{\text{irr}} = \Gamma W_{s,\text{irr}},\\
 I_{\text{rev}} &= \Gamma I_{s,\text{rev}}+ (1-\varrho)\mathbf{b}, &&
 \;\;I_{\text{irr}} = \Gamma I_{s,\text{irr}}+\varrho\mathbf{b},\notag
\end{align}
which ensures that they are tangent to the affine subspace, and thus neither component causes the process to drift away from it. {\color{black} Eq.~(\ref{eq: RNN HHD 1}) implies a decomposition of the RNN into reversible and irreversible components. In order to obtain these components, we write the projected dynamics in the affine subspace,
\begin{align}
\label{eq: RNN HHD irrev}
    \hat{\mathbf{f}}_{\text{irr}}(\mathbf{y}) &=-\varrho \mathbf{y + }W_{s,\text{irr}}h(\Gamma \mathbf{y} + \mathbf{b}) + I_{s,\text{irr}},\\
    \hat{\mathbf{f}}_{\text{rev}}(\mathbf{y}) & =-(1-\varrho)\mathbf{y}  +W_{s,\text{rev}}h(\Gamma \mathbf{y} + \mathbf{b}) + I_{s,\text{rev}},
\label{eq: RNN HHD rev}
\end{align}
where we notice that these can be seen as a pair of two-layer perceptrons, each approximating a component of the drift. However, the weights and biases of the first layer, $(\Gamma, \mathbf{b})$, are both shared and fixed. The \textit{sum} of the weights and biases in the second layer are also fixed i.e. $W_{s} = W_{s,\text{rev}} + W_{s,\text{irr}}$ and $I_{s} = I_{s,\text{rev}} + I_{s,\text{irr}}$. Finally, we have a trainable parameter $\varrho \in [0,1]$.

In order to obtain the components of the decomposition of the RNN, we must, again, train a two-layer perceptron, this time to approximate the irreversible component under these constraints: the weights and biases of the first layer are fixed; the `twin' perceptron defined by the same parameters in Eq.~(\ref{eq: RNN HHD rev}) approximates the reversible component. More explicitly, we introduce the HHD loss-function,
\begin{align}
\mathcal{L}_{\text{HHD}} &= ||\mathbf{f}_{\text{irr}}(\mathbf{y})+ \varrho\mathbf{y} - W_{s,\text{irr}}\mathbf{h}(\Gamma \mathbf{y}+ \mathbf{b}) - I_{s,\text{irr}}|| \\&+
||\mathbf{f}_{\text{rev}}(\mathbf{y})+ (1-\varrho)\mathbf{y} - (W_s-W_{s,\text{irr}})\mathbf{h}(\Gamma \mathbf{y}+ \mathbf{b}) - (I_s-I_{s,\text{irr}})||\notag,
\end{align}
where we optimise over $W_{s,\text{irr}}$, $I_{s,\text{irr}}$, and $\rho$; whilst $\{\Gamma, \mathbf{b}, W_s, I_s,B_s\}$ are fixed from training the original RNN.

Under these additional constraints, the two-layer perceptron is no longer guaranteed to be a universal approximator. However, in practice, we find that this parametrisation is able to approximate the components with very high accuracy. This is most likely due to the fact that $(\Gamma, \mathbf{b})$ are optimised for the full drift field and therefore remain well-aligned with this downstream `fine tuning'. This procedure yields a HHD-like decomposition of the network structure that we can compare to the decomposition in Sec.~\ref{sec: symmetric-asymmetric decomp}.}

\begin{figure*}
    \centering
    \includegraphics[width=\linewidth]{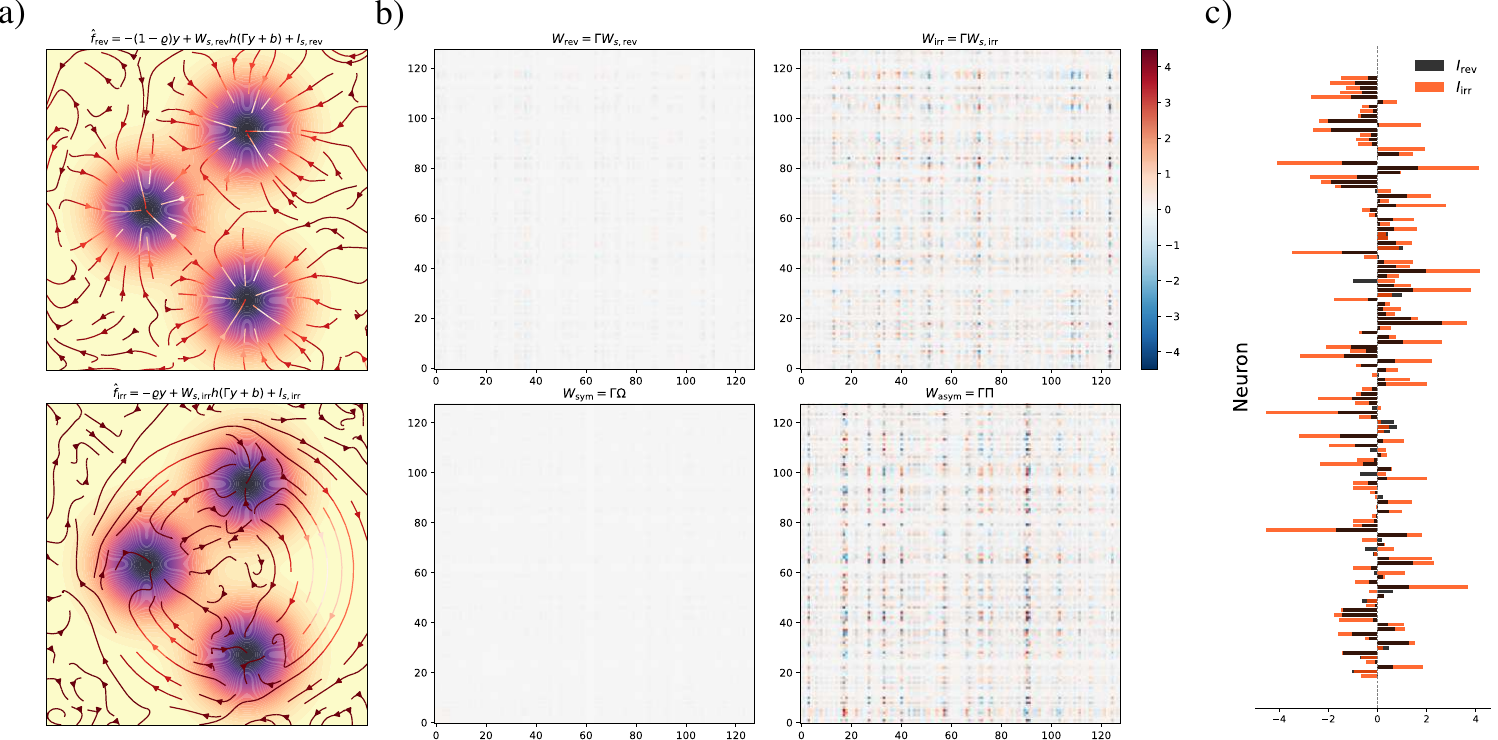}
    \caption{\textbf{The HHD of an RNN.} $a)$ Following Eq.~(\ref{eq: RNN HHD 1}), we train two coupled RNNs to approximate the reversible and irreversible components of the HHD respectively. Their projected drift shows that they were able to learn the components effectively. $b)$ We can decompose the learnt connectivity matrix, $W$, using both the HHD, $W = W_{\text{rev}} + W_{\text{irr}}$, or using the symmetric-asymmetric decomposition from Sec.~\ref{sec: symmetric-asymmetric decomp}, $W= W_{\text{sym}} + W_{\text{asym}}$. We find that for this three minima example, the magnitude of the irreversible and asymmetric components are far greater than the reversible component suggesting that encoding rotational or cycling dynamics requires more network {\color{black} asymmetry}. $c)$ We can also examine the decomposition of the input current vector, $I = I_{\text{irr}} + I_{\text{rev}}$. As with the connectivity, more of the input current is used to encode the irreversible component than the reversible component.}
    \label{fig: HHD}
\end{figure*}

Using the three minima example, we decompose the RNN with both the HHD and the symmetric-asymmetric decomposition. Panel $a)$ of Fig.~\ref{fig: HHD} shows the learnt HHD components of the vector field using Eq.~(\ref{eq: RNN HHD 1}), whilst Panels $b)$ and $c)$ show the decomposed connectivity matrix and bias vector, respectively. We find that for both decompositions, the irreversible/asymmetric component has significantly larger weights, in absolute value, hinting that irreversible cycling dynamics require far higher network complexity.

In the appendix, we further investigate how the character of the target dynamics is represented in the RNN structure. In App.~\ref{app: asymm EPR}, we show that the asymmetry of the RNN does not seem to be directly correlated with the EPR of the underlying process, a result that is in contrast with previous results for solvable networked systems \cite{nartallokaluarachchi2024broken}. In App.~\ref{app: chaos}, we describe how low-rank RNNs as a universal representation for nonlinear systems can be used to compare the complexity of dynamical systems, using chaotic attractors as an example.

\section{Conclusion}
\label{sec: conclusion}
{\color{black}
Neural circuits encode information and perform computations through a repertoire of complex spatiotemporal dynamics. Whilst their fundamental mechanisms remain obscure, RNNs have been shown to be a practical and plausible tool for studying neural dynamics both theoretically and empirically \cite{Vyas2020computation, Durstewitz2023reconstructing}. Neural data analysis has shown that high-dimensional brain network dynamics show significant redundancy, and that interpretable mechanisms can be observed at the level of a latent neural manifold \cite{Perich2025neuralmanifold,Gallego2017manifolds}. Many previous studies have attempted to bridge the theory of neural manifolds and the dynamics of RNNs by considering the implications of low-rank structure \cite{Mastrogiuseppe2018linking, Schmutz2025highdim, Schuessler2020dynamics, Schuessler2022interplay, Valente2022extracting}, including showing that such networks can approximate low-dimensional vector fields \cite{Arora2025efficient, Beiran2021shaping, Schuessler2020dynamics}. Extending this connection to stochastic dynamics, we presented a simple procedure to train a low-rank RNN to encode an arbitrary target SDE in an affine latent subspace, which allows us to encode nonlinear and nonequilibrium processes, such as chaotic attractors, and showed that RNNs are able to perform both input-driven switching and autonomous cycling between attracting states. Finally, in an effort to elucidate the representation of the dynamics that is learnt by the RNN, we presented two decompositions of the learnt connectivity matrix, using its asymmetry and its time-irreversibility respectively.

Our approach has a number of limitations that we want to outline. First, unlike biologically plausible training mechanisms, we trained the RNNs via backpropagation of error, leveraging that a low-rank RNN is equivalent to a two-layer perceptron. When considering the specific task of encoding a target SDE with known drift and additive noise, this training procedure is quick, easy, and stable. Nevertheless it does not apply with the same generality of domains as BPTT and is not biologically plausible, like the collection of methods designed for training RNNs via local or biological learning rules \cite{Murray2019local, Bellec2020asolution,Miconi2017biologically,Brunel1996hebbian,Sussillo2009chaotic, Ninou2025Curl, Rajan2016recurrent}. As a result, neural manifolds do not \textit{emerge} spontaneously due to self-organisation, e.g. through a biological learning rule. Instead, they are created by enforced low-rank structure in the connectivity, in line with a number of previous studies that focus on low-rank \cite{Mastrogiuseppe2018linking, Schmutz2025highdim, Schuessler2020dynamics, Schuessler2022interplay, Valente2022extracting} or otherwise constrained connectivity \cite{Pollock2020engineering,Yishai1995orientation, Bernardo2025shaping}}. Nevertheless, the mechanisms by which biological neural circuits \textit{learn} to encode complex dynamics over neural manifolds remain unclear. Elucidating these mechanisms would require biologically plausible learning rules, for example through local weight updates, which produce low-dimensional manifold structure in an emergent fashion, rather than through direct constraints, and are consistent with neural data. Drawing a link between network training and the empirical mechanisms of memory formation and consolidation would further solidify the relevance of such models for cognitive neuroscience \cite{Wang2010hippocampal}. {\color{black} Whilst encoding dynamics in RNNs is an interesting approach for understanding how computation may be performed in neural circuits, another relevant direction not considered here is parameter analysis of RNNs trained for more arbitrary computational tasks. Such an approach can also reveal latent structure that is emergent e.g. a line attractor representing a positive-negative spectrum in sentiment analysis \cite{Maheswaranathan2019reverse}. This can reveal how low-dimensional behaviour is an emergent feature of learning in a diversity of contexts.}

Next, our approach encodes dynamics in an affine linear subspace, assuming that the relationship between neuronal states and neural manifolds is linear. Whilst linear methods have been used extensively to probe neural data \cite{Cunningham2014dimensionality, Churchland2012neural}, state-of-the-art approaches reveal neural manifolds with nonlinear dimensionality reduction \cite{Perich2025neuralmanifold, Pandarinath2018lfads, Gosztolai2025interpretable,Gallego2017manifolds, Schneider2023cebra}. A more general relationship between neural firing and the effective computational manifold would allow for a much richer latent geometry in the system, which would be more consistent with empirical data. However, this generalisation leads to major mathematical challenges as nonlinear maps introduce difficult Jacobian and Hessian terms in the projected dynamics. {\color{black} As mentioned in Sec.~\ref{sec: target matching}, we also assume that the noise structure is both correlated and trainable rather than reflecting measured noise statistics from empirical data \cite{ebrahimi2022emergent}. It is as of yet unclear how a neural circuit could optimise these correlations in a biologically plausible fashion.} In order to build a closer relationship between our framework and modern approaches to neural manifolds, future work should focus on the inverse problem of reconstructing low-dimensional vector fields or operators from observed trajectories over neural manifolds \cite{brunton2022datadriven, brunton2022koopman}, {\color{black} in line with some recent approaches that have attempted to simultaneously learn both the manifold and its dynamics \cite{Gosztolai2025interpretable,Champion2019sindyautoencode,Muratore2026extracting}}. Furthermore, we focus on a `firing-rate model', which assumes that information in neural circuits is encoded in the instantaneous firing rates of individual neurons \cite{hopfield1984continuous}. This is in contrast with more biologically-plausible extensions, such as spiking neural networks, which have biophysically-consistent dynamics for each individual neuron \cite{Izhikevich2006dynamical} and typically assume that exact spike-times encode information. {\color{black} As mentioned in Sec.~\ref{sec: ddm}, we also assumed an additive noise structure with tailored correlations. Whilst this allowed us to tune the noise such that the dynamics on the latent manifold are isotropic, it is a significant restriction which presupposes that such an intricate tuning of noise is possible in real-world neural networks. Future work with nonlinear manifolds could attempt to alleviate this restriction by assuming isotropic noise at the level of the network, whilst achieving a target noise structure on the latent manifold through a nonlinear projection.} Finally, whilst we considered one case of an input-driven circuit, we mostly focused on autonomous RNN dynamics, which encode stationary processes. Integrating an RNN as the computational circuit in a more complete and time-dependent network, which includes both sensory input and an upstream output, is crucial for relating this model to more neuroscientific applications, such as motor control.

The ability of RNNs to encode complex, nonlinear dynamics can be interestingly related to the emerging topic of nonequilibrium brain dynamics. At a range of scales, the brain has been shown to violate the detailed balance condition and produce time-irreversible dynamics \cite{nartallokalu2025review}. Whilst nonequilibrium brain network dynamics have been previously linked to asymmetric connections between neurons and brain regions \cite{yan2013landscape,nartallokaluarachchi2024broken, aguilera2023sherrington, aguilera2021meanfield}, we go beyond observational analysis of neural data or the treatment of solvable models, and provide a framework that allows us to explore possible computational \textit{mechanisms} that require nonequilibrium dynamics. By encoding sequential information, in the form of irreversible SDEs, we suggest that time-irreversibility is an emergent feature of sequential computation in neural circuits. However, as discussed in App.~\ref{app: asymm EPR}, we found that the asymmetry of the RNN was not directly correlated with the irreversibility of the encoded process. Related works have shown conflicting results with respect to the correlation between the irreversibility of sensory stimuli and the irreversibility of the associated neural dynamics \cite{lynn2022decomposing, Aguilera2026inferring}, thus the exact relationship between the nonequilibrium nature of sensory information, computational mechanisms, and observed dynamics remains unclear.

In an effort to understand how dynamics are represented by an RNN, it is interesting to study the properties of the inferred connectivity. In particular, as any symmetric component of the connectivity can be seen as encoding an energy landscape, probing the connectivity with symmetric-asymmetric matrix decompositions is a promising way forward. Furthermore, a number of decompositions for stochastic processes have been developed \cite{DaCosta_2023, duong2023generic, Ma2015mcmc}, which have implications for their nonequilibrium thermodynamics and their convergence to stationarity. We present avenues for exploring the relationship between components of the connectivity, and the components of the underlying target process. 

{\color{black} In our work, we presented and exploited a simple approach for encoding low-dimensional dynamics with specified drift and diffusion coefficients in a high-dimensional, but low-rank, RNN. As we illustrated with crafted examples, it offers a theoretical framework for investigating the relationship between neural circuits and stochastic dynamics over neural manifolds, providing us with a way to investigate how low-dimensional cognitive processes may be performed.}

\section*{Code Availability}
Code is available at \url{https://github.com/rnartallo/embedding}.

\section*{Author contributions}
R.N.K. designed and performed research and wrote the manuscript. R.L. and A.G. designed and supervised the research and edited the manuscript.

\section*{Acknowledgements}
R.N.K would like to thank Pauline Thibaut for her pixel renditions of a dog and chick in App.~\ref{app: high dim cycling}. Additionally, the authors would like to thank photographer Alexas Fotos for uploading their image of the labyrinth game for free use (Fig.~\ref{fig: labyrinth}).  

R.N.K acknowledge support in the form of an EPSRC Doctoral Scholarship from Grants No. EP/T517811/1 and No. EP/R513295/1. R.L. acknowledges
support from the EPSRC grants EP/V013068/1, EP/V03474X/1 and EP/Y028872/1

\appendix
\section*{Supplementary Material}
\section{Numerical methods}

\subsection{Simulating trajectories from SDEs}
\label{app: sim}

We sample paths from an SDE using the \textit{Euler-Maruyama} discretisation,
\begin{align}
    \mathbf{x}_{t+\Delta t} = \mathbf{x}_t + \Delta t[\mathbf{f}(\mathbf{x}_t)] + \Delta \mathbf{w}_t[\Sigma(\mathbf{x}_t)], 
\end{align}
where $\Delta t$ is the time-step and $\Delta \mathbf{w}_t$ are independent and identically distributed (i.i.d.) multivariate normal random variables with mean zero and covariance $\Delta t \mathbf{I}_n$ \cite{kloeden1992sdes}.

\subsection{Computing inputs for attractor-switching}
\label{app: computing c}

In Sec.~\ref{sec: switching}, we propose an approach for designing an RNN that switches between attractors based on an input. Here we present the concrete approach used for computing $\mathbf{c}$, $G_s$, and $\mathbf{d}$, in the numerical example. As detailed in Sec.~\ref{sec: switching}, $\mathbf{c}\in \mathbb{R}^k$ makes the local minimum $\bm{\mu}_{i^*}$ a global minimum if,
\begin{align}
    \mathbf{c}\cdot(\bm{\mu}_{i^*}-\bm{\mu}_{j})>V^0_{i^*}-V^0_{j},
\end{align}
for all $j\neq i^*$. Whilst this defines a convex subset, we will focus on finding a single value $\mathbf{c}^*_{i}$ for each $i$ by solving the quadratic convex problem,
\begin{align}
\mathbf{c}^*_i&=\text{argmin}_{\mathbf{c}\in\mathbb{R}^k}\frac{1}{2}||\mathbf{c}||_2^2,\\
    \text{s. t. } \;\mathbf{c}\cdot(\bm{\mu}_{i}-\bm{\mu}_{j})&\geq V^0_{i}-V^0_{j} + \delta, \text{ for } j\neq i,
\end{align}
where $\delta>0$ is a small parameter used to weaken the strict inequality. We solve this with convex optimisation in \texttt{cvxpy}. Given a collection of $\{\mathbf{c}_j^*\}$, we choose a coding for the inputs. For example, we can choose a `one-hot' encoding where $l=m$ i.e. we have one input direction per minimum, and $\mathbf{s}=\mathbf{e}_{i}$ is the input corresponding to minimum $i$, where $\mathbf{e}_{i}$ is the unit vector in the direction $i$. Alternatively, we can sample random Gaussian vectors for $\mathbf{s}$. In any case, we can also add low-intensity noise to the input. It remains to solve for $G_s$ and $\mathbf{d}$ given the coding scheme, for which we use \textit{ridge regression} \cite{Hastie2009elements}. We combine our $\mathbf{c}^*$ values into a matrix $\mathbf{C}\in \mathbb{R}^{m\times k}$, and combine the codes $\mathbf{s}_i$ into a matrix $\mathbf{S}\in \mathbb{R}^{m\times m}$, and augment it with an additional column of ones to form $\mathbf{S}^* = [\mathbf{S}, \mathbf{1}]\in \mathbb{R}^{m\times (m+1)}$. We then aim to fit the model $\mathbf{C} \approx \mathbf{S}\mathbf{G_s} + \mathbf{1}\mathbf{d}^{\top}$, by minimising,
\begin{align}
    \Theta^* & = \text{argmin}_{\Theta}||\mathbf{S}^*\Theta - \mathbf{C}||^2 + \lambda||\Theta||^2,
\end{align}
where $\Theta = [G_s, \mathbf{d}]$. This has exact solution,
\begin{align}
    \Theta^*= ((\mathbf{S}^*)^{\top}\mathbf{S}^*+\lambda \mathbf{I})^{-1}(\mathbf{S}^*)^{\top}\mathbf{C}.
\end{align}

\begin{figure*}
    \centering
    \includegraphics[width=\linewidth]{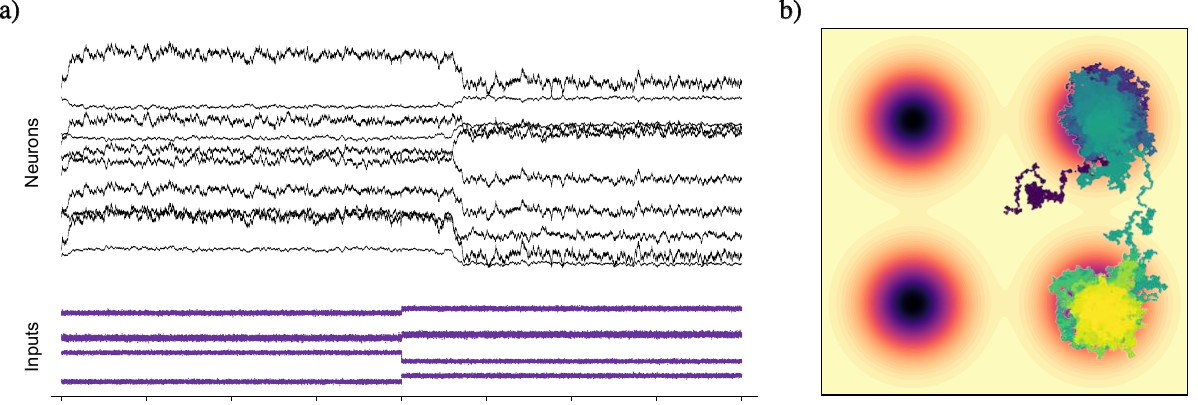}
    \caption{\textbf{Input-driven attractor switching with random Gaussian codes.} $a)$ Instead of a `one-hot' encoding, we can also take random Gaussian vectors to represent each memory. We add Gaussian noise to this input signal and repeat the experiment from before, switching the input signal from minimum 1 to minimum 2 halfway along the trajectory. $b)$ As the input noise intensity is smaller than the difference between the codes, the network stills switches between the `memories' correctly, which can be seen at the level of the neural manifold.}
    \label{fig: random Gaussian codes}
\end{figure*}

{\color{black} In the main text, we consider a `one-hot' coding scheme. In Fig.~\ref{fig: random Gaussian codes}, we perform the same experiments but with random Gaussian codes and a signal that is subject to Gaussian noise. When the noise intensity is low enough that the network can differentiate between codes, this coding scheme works effectively as shown by the attractor switching in Fig.~\ref{fig: random Gaussian codes}.}

\subsection{Constructing closed curves with splines}
\label{app: splines}
We compute interpolating closed curves with cubic splines with periodic boundary conditions. More specifically we use \texttt{CubicSpline} from \texttt{scipy.interpolate} \cite{Virtanen2020scipy}. We then parametrise the curve, $\gamma(t)$, with 500 time-points and construct a \texttt{matplotlib} path. We can then measure the distance between any point, $p$, and each discrete point in the path, taking the minimum to be the distance between $p$ and $\gamma$.

{\color{black}
\section{Background: Hopfield's model}
\label{app: hopfield}

In his seminal paper \cite{hopfield1984continuous}, Hopfield considered the RNN in Eq.~(\ref{eq: RNN}) with the connectivity constrained to be symmetric, $W= W^{\top}$. He introduced the \textit{energy potential},
\begin{align}
\label{eq: energy}
    E(\mathbf{v}) = -\frac{1}{2}\sum_{j,i}^nW_{ij}v_{i}v_{j} - \sum_{i}I_iv_i + \sum_{i}^n\int^{v_i}_0h^{-1}(\omega) \;d\omega,
\end{align}
and $\tilde{E}(\mathbf{u}) = E(\mathbf{h}(\mathbf{u}))$, where the drift of the RNN can be written in the form,
\begin{align}
    \mathbf{F}(\mathbf{u}) & =  -\nabla_{\mathbf{v}}E(\mathbf{v})=- \mathbf{D}(\mathbf{u})\nabla_{\mathbf{u}}\tilde{E}(\mathbf{u}),
\end{align}
where $\mathbf{D}(\mathbf{u}) = \text{diag}\left(\frac{1}{h'(u_1)},...,\frac{1}{h'(u_n)}\right)$ is a positive-definite matrix-valued function. As a result, the RNN is constrained to perform `generalised' gradient descent on the energy landscape, eventually arriving at an energy minimum.\footnote{The prefactor $\mathbf{D}(\mathbf{u})$ causes the gradient to be scaled unevenly in different directions, thus the system does not follow the Euclidean direction of steepest descent on the landscape. Instead, the drift is a gradient flow with respect to the Riemannian metric defined by $\mathbf{D}(\mathbf{u})^{-1}$. Practically, we would see that the system descends the landscape monotonically, but more quickly in directions where $h'(u_i)$ is small.} 

Hopfield's model allows for the manipulation of the energy landscape via training, such that a desired `memory' can be encoded as an energy minimum. The result is an autonomous neural circuit that will `retrieve the memory given initial information' i.e. converge to the energy minimum from an appropriate initial state. Hopfield's results laid the ground work for the study of \textit{attractor neural networks} and their possible implications for neuroscience \cite{Amit1989modelingbrain, Khona2022attractor}.
}

\section{Optimal decomposition of a network under constraints}
\label{app: decomp}

Let $W \in \mathbb{R}^{n \times n}$ be a matrix that is low-rank and has the form $W= \Gamma W_s$, where $\Gamma \in \mathbb{R}^{ n \times k}$ and $W_s \in \mathbb{R}^{k \times n}$. $\Gamma$ has full column rank $k$. We want to find the `best' decomposition,
\begin{align}
W = \Gamma(\Omega+\Pi),
\end{align}
where $\Gamma \Omega$ is symmetric. We will define this optimal decomposition to have,
\begin{align}
    \tilde{\Omega} = \text{argmin}_\Omega||\Gamma \Omega-C||_F^2,
\end{align}
where $C = \frac{1}{2}\left(W + W^{\top}\right)$ i.e. the matrix that is most similar to the symmetrised version of $W$, but which respects the parametrisation.

Let $P$ be the projector matrix onto
\begin{align}
    \mathcal{S} = \{X \in \mathbb{R}^{n \times n}: X \in \text{range}(\Gamma)\},
\end{align}
i.e. $P = \Gamma \Gamma^{\dagger}$. Due to the symmetry of any feasible $X$, we have that both the columns and rows of $X$ live in the range of $\Gamma$. This implies that,
\begin{align}
    PX = X,&& XP = X,
\end{align}
which implies that $X = PXP$ for any feasible $X$.

As we are trying to minimise the Frobenius norm, the optimal solution $X = \Gamma \Omega$, is the projection of $C$ onto the linear subspace,
\begin{align}
    \mathcal{Q} = \{X \in \mathbb{R}^{n \times n}: X=X^{\top}, X \in \text{range}(\Gamma)\}.
\end{align}
As $C$ is already symmetric, the solution is simply the projection of $C$ onto $\mathcal{S}$,
\begin{align}
X = PCP = \Gamma \Gamma^{\dagger}C\Gamma \Gamma^{\dagger},
\end{align}
which gives,
\begin{align}
    \tilde{\Omega} = \Gamma^{\dagger}C\Gamma \Gamma^{\dagger}.
\end{align}

{\color{black}\section{Training via linear regression: the neural engineering framework}
\label{app: regression}

An alternative approach to training low-rank RNNs to approximate {\color{black} vector fields} is to consider the RNN as defining a collection of random nonlinear basis functions, whose coefficients can be inferred via linear (or ridge) regression -- this is precisely the set-up considered in the NEF \cite{Eliasmith2003neuralengineering, Arora2025efficient}.

Whilst we choose to use backpropagation of error to learn both the network and the subspace, a simpler, albeit less precise, approach is to sample $(\Gamma, \mathbf{b})$ from a random distribution, which are then fixed, and then solve for $W_s$ and $I_s$ using linear regression.

\begin{figure}
    \centering
\includegraphics[width=\linewidth]{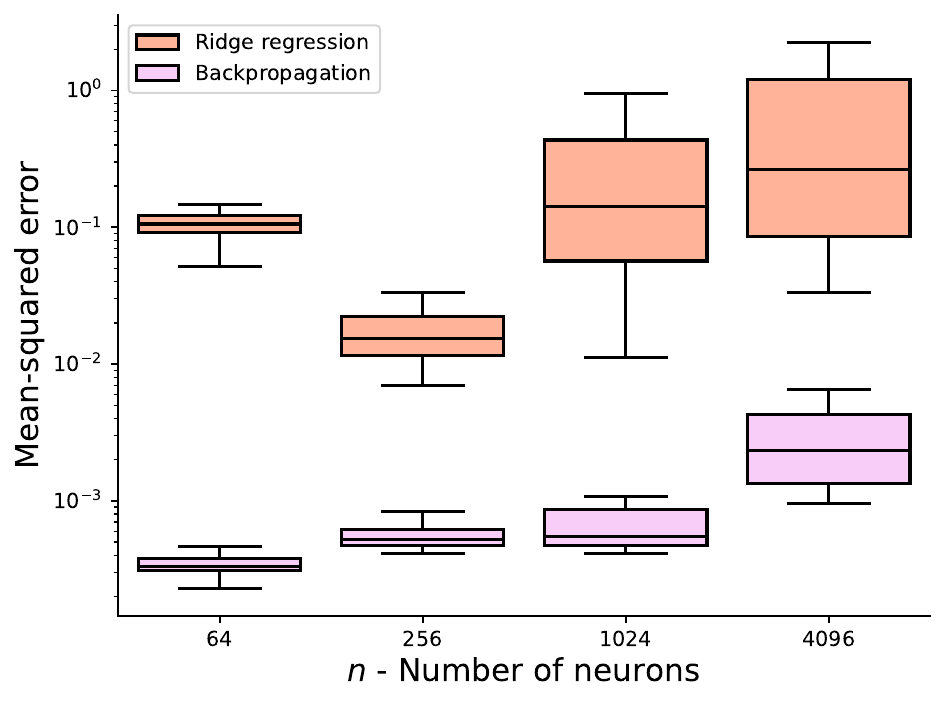}
    \caption{\textbf{Backpropagation vs. ridge regression.} We compare the mean-squared error of RNNs trained to approximate the VDP system using either backpropagation or ridge regression, across a range of network sizes and repeated simulations. We find that backpropagation significantly outperforms ridge regression.}
    \label{fig: reg_vs_backprop}
\end{figure}

To compare this with our backpropagation approach, we considered the VDP system and generated a single set of training data. Next, we fit RNNs of sizes 64, 256, 1024, and 4096 neurons using backpropagation and ridge regression, with 50 simulations at each network size. Fig.~\ref{fig: reg_vs_backprop} shows network size against mean-squared error at the sampled points. We found that backpropagation significantly outperforms ridge regression, as expected due to the increased representational complexity allowed by learning both the subspace and the network parameters simultaneously. Nevertheless, regression is a simple and quick alternative approach that has a closed form solution (see App.~\ref{app: sim}.2). For these simulations we set $\lambda_{\text{diff}} = 0$ to focus only on fitting the drift field.

Previous work has found that in an RNN trained via a feed-back signal from a linear read-out, recursive regression outperforms both FORCE learning and BPTT \cite{Arora2025efficient}.
}

{\color{black}\section{Nonlinear examples}
\label{app: nonlinear examples}
In Sec.~\ref{sec: examples 1} we consider the following nonlinear systems:
\begin{itemize}
\item
The stochastic \textit{van der Pol} (VDP) system is a nonlinear, planar system with a stable limit cycle. It is given by the system,
\begin{align}
     d\begin{pmatrix}
y_1(t) \\y_2(t)
\end{pmatrix} = \begin{pmatrix}
 y_2\\-y_1+\mu y_2(1-y_1^2)
\end{pmatrix}\;dt + \sigma d\begin{pmatrix}
 w_1(t)\\w_2(t)
\end{pmatrix},
\end{align}
where we take $\mu = 1$ \cite{Sprott2010chaos}.

\item The stochastic \textit{Lorenz system} (LS) is a three-dimensional chaotic system given by,
\begin{align}
     d\begin{pmatrix}
y_1(t) \\y_2(t)\\y_3(t)
\end{pmatrix} = \begin{pmatrix}
 \varsigma(y_2- y_1)\\y_1(\rho-y_3)-y_2\\
 y_1y_2-\beta y_3
\end{pmatrix}\;dt + \sigma d\begin{pmatrix}
 w_1(t)\\w_2(t)\\w_3(t)
\end{pmatrix}.
\end{align}
The system has a characteristic strange attractor with two scrolls when $\varsigma = 10$, $\rho = 28$, and $\beta = 8/3$ \cite{Sprott2010chaos}.

\item
The stochastic \textit{Dadras system} (DS) is a three-dimensional chaotic system given by,
\begin{align}
     d\begin{pmatrix}
y_1(t) \\y_2(t)\\y_3(t)
\end{pmatrix} = \begin{pmatrix}
 y_2-py_1+oy_2y_3\\ry_2-y_1y_3+y_3\\
 cy_1y_2-ey_3
\end{pmatrix}\;dt + \sigma d\begin{pmatrix}
 w_1(t)\\w_2(t)\\w_3(t)
\end{pmatrix}.
\end{align}
The system has a strange attractor with three scrolls when $p=3$, $o=2.7$, $r=1.7$, $c=2$, and $e=9$ \cite{Dadras2009chaos}.
\end{itemize}
}

{\color{black}\section{Reconstructing affine subspaces from trajectories}
\label{sec: reconstructing}
Another pertinent question is whether or not we can reconstruct the appropriate affine subspace given only trajectories from the RNN. Given dynamics that are entirely constrained to an affine subspace, it is possible to exactly identify the subspace, up to numerical accuracy. The solution comes from a seminal result by Pearson, and is known as \textit{principal component analysis} (PCA) \cite{Pearson1901pca}. If the trajectory is not perfectly constrained i.e. there is isotropic Gaussian variation in the directions orthogonal to the subspace, then an exact recovery is not possible, but PCA becomes the maximum-likelihood estimator of the subspace. Subspace identification through PCA is extremely common in the analysis of neural data, and has been used in countless studies to identify neural manifolds in brain network activity \cite{Churchland2012neural,Mitchell2023neuralmanifold,Cunningham2014dimensionality}, as well as to identify latent subspaces in the activity of trained RNNs \cite{Sussillo2013opening,Cueva2021rnn}.

This method is canonical and the proofs have been reviewed at length (e.g. \cite{Vidal2016generalisedpca}), thus we will only outline the approach in brief, and illustrate how it applies to our setting.
Given trajectory data of the form $\{\mathbf{u}(t_j): j = 1,...,T\}$, we assume the data lives in an affine subspace of dimension $k$, where $k$ typically begins as an unknown, and can be written in the form,
\begin{align}
    \mathbf{u} = \Psi \bm{\upsilon} + \bm{\mu},
\end{align}
where $\Psi \in \mathbb{R}^{n\times k}$ has orthonormal columns, $\bm{\mu}\in \mathbb{R}^n$, and $\bm{\upsilon}(t)\in\mathbb{R}^k$ is a parametrisation.

\begin{figure*}
    \centering
    \includegraphics[width=\linewidth]{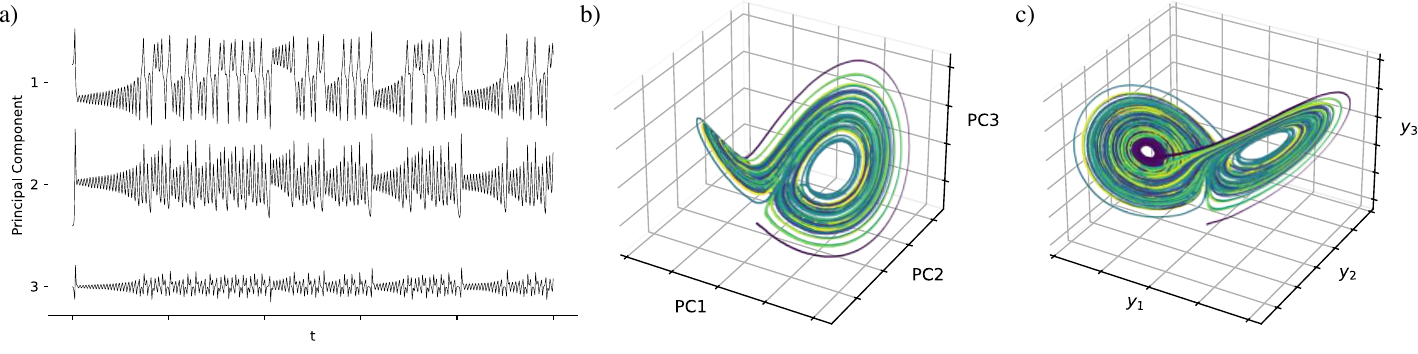}
    \caption{\textbf{Reconstructing latent subspaces from trajectories}. $a)$ Trajectories from a low-rank RNN trained to embed a $k-$dimensional system only have $k$ non-zero principal components. $b)$ With PCA, we can identify the affine subspace in which the dynamics are constrained. $c)$ The PCA representation is equivalent to original coordinates up to an affine linear transformation. Given $(\Gamma, \mathbf{b})$ can solve for this transformation to recover the original coordinates.}
    \label{fig: pca}
\end{figure*}

Whilst it is tempting to associate $\Psi$ with $\Gamma$ and $\bm{\mu}$ with $\mathbf{b}$, we note that this is not necessarily the case. Whilst the subspace can be identified, the exact parametrisation cannot be, as for any invertible matrix $A$, we have indistinguishable parametrisations,
\begin{align}
    \mathbf{u} &= \Gamma \mathbf{y} + \mathbf{b}= \underbrace{(\Gamma A^{-1})}_{\widetilde{\Gamma}}\underbrace{(A(\mathbf{y}- \mathbf{c}))}_{\mathbf{\widetilde{y}}} + \underbrace{\mathbf{b} +\Gamma \mathbf{c}}_{\mathbf{\widetilde{b}}}= \widetilde{\Gamma} \mathbf{\widetilde{y}} + \mathbf{\widetilde{b}}.
\end{align}
Thus, the low-dimensional trajectory we obtain is an affine transformation of the original parametrisation.

We stack our time-series into the matrix $\mathbf{U} \in \mathbb{R}^{n \times T}$ and center it, $\mathbf{X} = \mathbf{U} - \bar{\mathbf{u}}$, where
\begin{align}
    \bar{\mathbf{u}} = \frac{1}{T}\left (\sum_{j}u_1(t_j),...,\sum_{j}u_n(t_j)\right).
\end{align}
This is equivalent to choosing $\bm{\mu}=\bm{0}$. Next, we take the singular value decomposition (SVD) of $\mathbf{X}$. For long trajectories, it is more efficient to compute the covariance matrix, $\Theta = \mathbf{X}\mathbf{X}^{\top}$, and compute its eigendecomposition. Assuming that the dynamics are exactly constrained to the subspace, then this matrix will have $k$ non-zero eigenvalues, whose eigenvectors form the columns of $\Psi$, which we call \textit{principal components} (PCs), and where we order the columns by decreasing eigenvalue magnitude, $\{\lambda_1,...,\lambda_k\}$ and $\lambda_j = 0$ for $j \in \{k+1,...,n\}$.\footnote{The covariance matrix is symmetric positive semi-definite and thus has only non-zero eigenvalues.} The zero eigenvalues correspond to the directions orthogonal to the subspace in which the dynamics do not evolve. For noisy data, these eigenvalues are non-zero but very small. In this case, the dimension of the subspace, $k$, can be approximately identified by considering the \textit{proportion of the variance explained} by each PC, $\vartheta_l = \lambda_l/\sum_{j=1}^n\lambda_j$, where we `cut-off' after $k$ components, if $\sum_{j=1}^k \vartheta_j> c$, where $c$ is a threshold, e.g. $c=0.7$ \cite{Vidal2016generalisedpca}. We then project the time-series to obtain the parametrisation, $\bm{\upsilon} = \Psi^{\top}\mathbf{X}$. Panels $a-b)$ of Fig.~\ref{fig: pca} show the traces and phase-space dynamics of the 3 non-trivial PCs for a trajectory from the 512-neuron RNN trained to embed the LS.

Given $\Gamma$ and $\mathbf{b}$, we can find the affine transformation that turns $\bm{\upsilon}(t)$ to $\mathbf{y}(t)$. We can solve $\Gamma \Xi = \Psi$ and $\Gamma \bm{\chi}  = \bar{\mathbf{u}} + \mathbf{b}$ as a least-squares problem, and apply the transformation $\mathbf{y}(t) = \Xi \bm{\upsilon}(t) + \bm{\chi}$, to obtain the original parametrisation, as shown in Panel $c)$ of Fig.~\ref{fig: pca}.}

\section{Designed rotational drift is DFSD}
\label{app: coplanar DFSD}

In the example considered in Sec.~\ref{sec: attractor design}, we consider a rotational field of the form,
\begin{align}
    \mathbf{R}(\mathbf{y}) & = \frac{\alpha}{\pi(\mathbf{y)}}g(\mathbf{y})\mathbf{M}\nabla\phi(\mathbf{y}).
\end{align}
We can show that this field is DFSD by showing that,
\begin{align}
    \bm{\zeta}(\mathbf{y}) & = g(\mathbf{y})\mathbf{M}\nabla\phi(\mathbf{y}),
\end{align}
is divergence-free. Applying the product rule we have that,
\begin{align}
    \nabla \cdot \bm{\zeta} & = (\nabla g)\cdot(\mathbf{M}\nabla\phi) + g(\nabla\cdot(\mathbf{M}\nabla \phi)).
\end{align}
We have that,
\begin{align}
    \nabla g & = -\frac{\phi}{\xi^2}g\nabla \phi,
\end{align}
which implies that,
\begin{align}
    (\nabla g)\cdot(\mathbf{M}\nabla\phi)& = -\frac{\phi}{\xi^2}g\nabla \phi\cdot(\mathbf{M}\nabla\phi),\\
    & = 0,\notag
\end{align}
as $\mathbf{v}\cdot\mathbf{M}\mathbf{v} = 0$ for antisymmetric $\mathbf{M}$. Next, we have that,
\begin{align}
    \nabla\cdot(\mathbf{M}\nabla \phi) & = \sum_{i,j}\mathbf{M}_{ij}\partial_i\partial_j\phi,
\end{align}
which is the sum-product of a symmetric (Hessian) and antisymmetric matrix, which is zero. Thus $\nabla \cdot \bm{\zeta} = 0$.

\section{Attractor-cycling diffusions in higher-dimensions}
\label{app: high dim cycling}

In this section, we consider similar approaches to that of Sec.~\ref{sec: attractor design} for designing SDEs with transiently visited attractors. First, we consider a high-dimensional example that cycles between two pixelated images. Next, we consider a three-dimensional example where the attracting points are non-coplanar. Unlike the examples in Sec.~\ref{sec: attractor design}, we do not encode these processes in the dynamics of an RNN as it is not motivated to encode a high-dimensional process on a neural manifold, and the non-coplanar example is not smooth.

\subsection{A high-dimensional example with pixel images}
\begin{figure*}
    \centering
    \includegraphics[width=\linewidth]{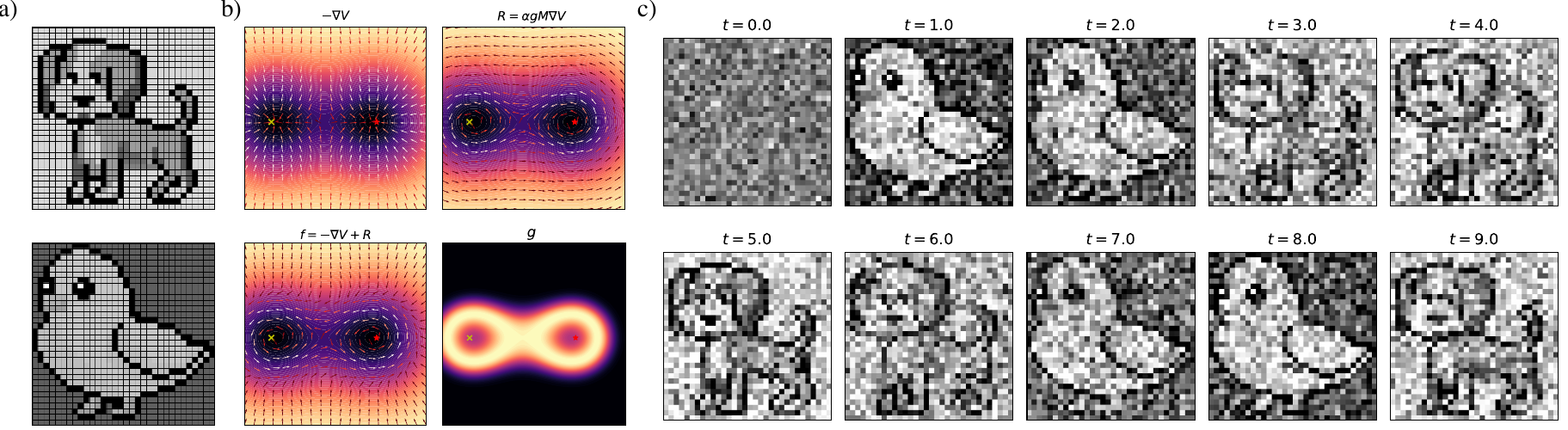}
    \caption{\textbf{Cycling between pixelated images in high-dimensions}. We can modify our approach from Sec.~\ref{sec: attractor design} to design processes which cycle between energy minima in high-dimensions. $a)$ We illustrate this using $32 \times 32$ pixelated images of a dog and a chick, which are points in $\mathbb{R}^{1024}$. $b)$ As before, we design a drift field that encodes these points as the minima of a Gaussian potential, but drives rotation between them without disrupting the stationary density. $c)$ Integrating this SDE, we can sample paths which cycle between (noisy) instances of the two images in the high-dimensional space.}
    \label{fig: pixels}
\end{figure*}
We consider two grey-scaled, $32 \times 32$ images, whose pixel-values are scaled to $[0,1]$. Panel $a)$ of Fig.~\ref{fig: pixels} shows a dog and chick image, respectively. These points can be thought of as points $\bm{\mu}_1,\bm{\mu}_2 \in  \mathbb{R}^{1024}$. We define a Gaussian potential on $\mathbb{R}^{1024}$ with minima at the points corresponding to the dog and chick images respectively. Next, we want to create a rotational flow that cycles between these pixelated images in $\mathbb{R}^{1024}$. To do this, we first select a plane which intersects both these points -- we choose the one that goes through the origin -- which is spanned by vectors $\{\mathbf{e}_1, \mathbf{e}_2\}$, using the Gram-Schmidt orthogonalisation. We take $\mathbf{d} = \bm{\mu}_2-\bm{\mu}_1$, and $\mathbf{e}_1 = \mathbf{d}/||\mathbf{d}||$. Next, we must choose a vector which is linearly independent of $\mathbf{e}_1$, so we choose the standard basis vector in the direction where $\mathbf{e}_1$ has smallest absolute value, $\bm{\epsilon}_k$, and then project this vector onto $\mathbf{e}_1$,
\begin{align}
    \mathbf{q} = \bm{\epsilon}_k - (\bm{\epsilon}_k\cdot \mathbf{e}_1)\mathbf{e}_1,
\end{align}
and then normalise it to obtain $\mathbf{e}_2 = \mathbf{q}/||\mathbf{q}||$. The result is an orthonormal basis, $\{\mathbf{e}_1, \mathbf{e}_2\}$, for a 2D plane that intersects both minima.

Given this plane, the matrix representing a $\pi/2$ rotation in the plane, is given by,
\begin{align}
    \mathbf{M} = \mathbf{e}_2\otimes \mathbf{e}_1 - \mathbf{e}_1\otimes \mathbf{e}_2,
\end{align}
where $\otimes$ is the outer product. We define the rotational component to be,
\begin{align}
    \mathbf{R}(\mathbf{y}) = \alpha g(\mathbf{y}) \mathbf{M} \nabla V(\mathbf{y}),
\end{align}
where we are rotating the gradient field itself. Note that we have not defined an intersecting curve as in Sec.~\ref{sec: attractor design}. Instead, we modify the weighting kernel,
\begin{align}
    g(\mathbf{y}) & = \exp\left(-\frac{(V(\mathbf{y}) - s)^2}{2\xi^2}\right),
\end{align}
such that it decays with the distance from a particular level set of the potential function with value $s$, rather than the distance from a curve. As a result, the rotational field creates flow along this particular level set, pushing trajectories between the minima. It is easy to show that this is DFSD. Finally, we define our combined drift field to be,
\begin{align}
    \mathbf{f} = -\nabla V + \mathbf{R},
\end{align}
thus it admits a stationary Boltzmann distribution. The gradient flow, rotational component, weighting kernel, and combined field are projected onto the plane in Panel $b)$ of Fig.~\ref{fig: pixels}. Integrating the SDE with this drift field, and suitably chosen noisy intensity, defines a process whose trajectories cycle between (noisy) instances of the two images in succession, as shown in Panel $c)$ of Fig.~\ref{fig: pixels}.

\subsection{Non-coplanar minima in three dimensions}
\begin{figure*}
    \centering
    \includegraphics[width=\linewidth]{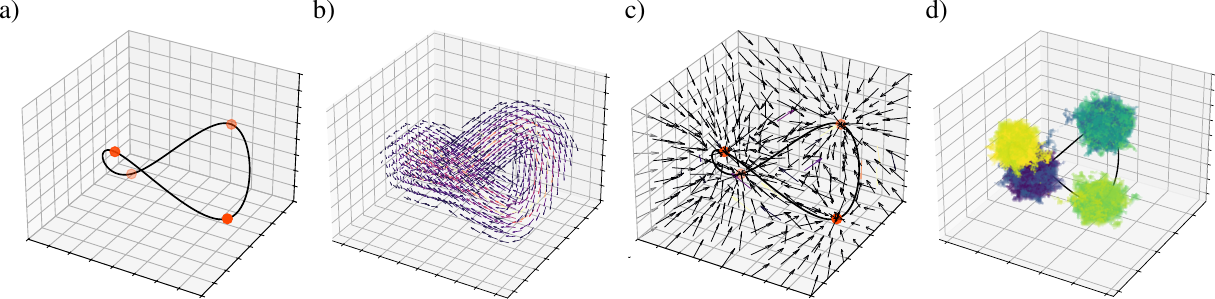}
    \caption{\textbf{Cycling between non-coplanar minima.} We can also design SDEs which cycle between non-coplanar minima in 3D. $a)$ In this example, we consider four minima on a saddle, which we interpolate with a cubic spline. $b)$ We design a rotational field which pushes trajectories around this curve, but is DFSD. $c)$ This produces a combined field which has both attracting and rotational dynamics. $d)$ Sampling from this process, we see that it rotates transiently between attracting states.}
    \label{fig: noncoplanar}
\end{figure*}
Here we consider a set of minima which are non-coplanar in $\mathbb{R}^{3}$. As before, we define the attractor dynamics, $-\nabla V$, with Gaussian wells. However, the rotational component $\mathbf{R}$ from the coplanar case cannot be applied in this setting. This is because the level sets of the signed distance function are no longer curves, but instead tubular sections. As a result, flows along the contours of these level sets are not restricted to the `direction' of the curve through the minima, but also along the minor radius of each tube. Instead, we introduce an alternative approach.

Again, we draw a differentiable curve, $\gamma$, that intersects our minima in the order that we wish them to be traversed. Fig.~\ref{fig: noncoplanar} shows an example with four minima which lie on saddle. Panel $a)$ shows the intersecting curve. Next, we consider the \textit{tubular neighbourhood}, $Q$, around the non-coplanar curve with parametrisation $\gamma(s)$, which is defined as,
\begin{align}
    Q = \{\mathbf{y}\in\mathbb{R}^{3}: \text{dist}(\mathbf{y},\gamma) < r\},
\end{align}
where $r$ is the \textit{tubular radius}, which must be smaller than the \textit{injectivity radius} $r_{\max} = \kappa_{\max}^{-1}$, where $\kappa_{\max}$ is the maximum curvature of $\gamma$ – this is the largest tube that does not self-intersect. Each point $\mathbf{y} \in Q$ can be written as,
\begin{align}
    \mathbf{y}(s,\rho,\theta) & = \gamma(s) + \rho\cos(\theta)\cdot\mathbf{N}(s) + \rho\sin(\theta)\cdot\mathbf{B}(s),
\end{align}
where $s$ is the arc-length parameter, and $(\rho,\theta)$ are polar coordinates in the \textit{normal plane} spanned by the \textit{normal} and \textit{bi-normal} vectors, $\{\mathbf{N}(s), \mathbf{B}(s)\}$ at $\gamma(s)$ \cite{Carmo1976differentiable,Lee2000smooth}. At any point $\gamma(s)$, we can form a local cylindrical coordinate system,
\begin{align}
    \mathbf{e}_s & = \mathbf{T}(s)\\\mathbf{e}_{\rho}&= \cos(\theta) \cdot \mathbf{N}(s) + \sin(\theta) \cdot \mathbf{B}(s)\notag\\
    \mathbf{e}_{\theta}&=-\sin(\theta)\cdot \mathbf{N}(s)+\cos(\theta)\cdot \mathbf{B}(s)  \notag
\end{align}
which are the \textit{axial}, \textit{radial} and \textit{azimuthal} basis vectors respectively, and $\mathbf{T}(s)$ is the \textit{tangent} vector to $\gamma(s)$ \cite{Carmo1976differentiable}. Together, the vectors $\{\mathbf{T}(s),\mathbf{N}(s), \mathbf{B}(s)\}$ form the \textit{Frenet-Serret} orthonormal frame along a curve \cite{Carmo1976differentiable}. This cylindrical basis is accurate up to a first-order correction arising due to the curvature of the curve. Our goal is to construct a vector field with forcing in the direction $\mathbf{T}(s)$. To do so, we define the vector potential,
\begin{align}
    \mathbf{A}(s,\rho,\theta) & = \psi(\rho)\mathbf{e}_{\theta}(s,\rho,\theta),
\end{align}
where $\psi(\rho) = \exp(-\rho^2/2\xi^2)$ is the weighting kernel that decays as $\rho$ increases i.e. that decays with distance from $\gamma$. This coincides with our definition of $g$ in the planar case (Eq.~(\ref{eq: weighting kernel})). We can take the curl to obtain a divergence-free field, $\nabla \times \mathbf{A}(\mathbf{x)}$, which, following the right hand rule, pushes in the direction $\mathbf{T}(s)$. In practice, the curl of the vector potential is approximated numerically using finite differences. Finally, dividing through by the stationary density, we obtain,
\begin{align}
    \mathbf{R}(\mathbf{y}) & = \frac{\alpha}{\pi(\mathbf{y})}\nabla \times \mathbf{A}(\mathbf{y}),
    \label{eq: non coplanar rotation}
\end{align}
where $\alpha$ controls the strength of rotation. This field is DFSD, thus $-\nabla V + \mathbf{R}$ converges to the Boltzmann density, yet pushes trajectories along $\gamma$, as desired. Panels $b)$ and $c)$ of Fig.~\ref{fig: noncoplanar} show the rotational component and combined vector field respectively, whilst Panel $d)$ shows an example trajectory from this process. It cycles autonomously between the attractors exhibiting metastability. As there is a discontinuity between the field inside the tubular radius and beyond it, the field is not smooth, thus we do not attempt to encode it in an RNN.

\section{Network asymmetry and entropy production}
\label{app: asymm EPR}

Neural systems operate in NESSs across a range of scales, with the level of irreversibility, defined by the EPR in Eq.~(\ref{eq: EPR}), being correlated with the level of consciousness or the complexity of a cognitive task \cite{nartallokalu2025review}. Moreover, the EPR has been shown to be driven by the asymmetry of the network underlying the neural activity \cite{nartallokaluarachchi2024broken}. Here, we investigate this further by considering a nonequilibrium diffusion with a parametrisable EPR, embedding the process in a latent subspace of a fixed-size RNN, and measuring the asymmetry of the resulting connectivity.

In particular, we consider the stochastic \textit{Hopf oscillator}, a nonlinear model with a solvable NESS, which is given by,
\begin{align}d
\begin{pmatrix}
    y_1(t)\\
    y_2(t)
\end{pmatrix}&= \begin{pmatrix}
    (1-y_1^2-y_2^2)y_1-\omega y_2\\
    (1-y_1^2-y_2^2)y_2+\omega y_1
\end{pmatrix}dt+ \sigma \begin{pmatrix}
    w_1(t)\\
    w_2(t)
\end{pmatrix},
\end{align}
where $\omega$ is the frequency of oscillation \cite{nartallokalu2025review}. The NESS is given by a radially-symmetric stationary density and, for fixed $\sigma$, has EPR $\Phi \propto \omega^2$, satisfying time-reversibility if and only if $\omega = 0$ (see Ref. \cite{Nartallo2025coarse}).

To measure the asymmetry of the resulting RNN, we consider the canonical symmetric-antisymmetric decomposition, the low-rank symmetric-asymmetric decomposition of Sec.~\ref{sec: symmetric-asymmetric decomp}, and the HHD of Sec.~\ref{sec: network hhd}. In general, this produces a pair $(W_1,W_2)$ where $W= W_1+W_2$, where we quantify the relative energy of $W_1$ with,
\begin{align}
\label{eq: network energy}
\mathcal{E}(W_1,W_2) &= \frac{|W_1|}{|W_1|+|W_2|},
\end{align}
where $|\cdot|$ is the Frobenius norm.
\begin{figure*}
    \centering
\includegraphics[width=0.85\linewidth]{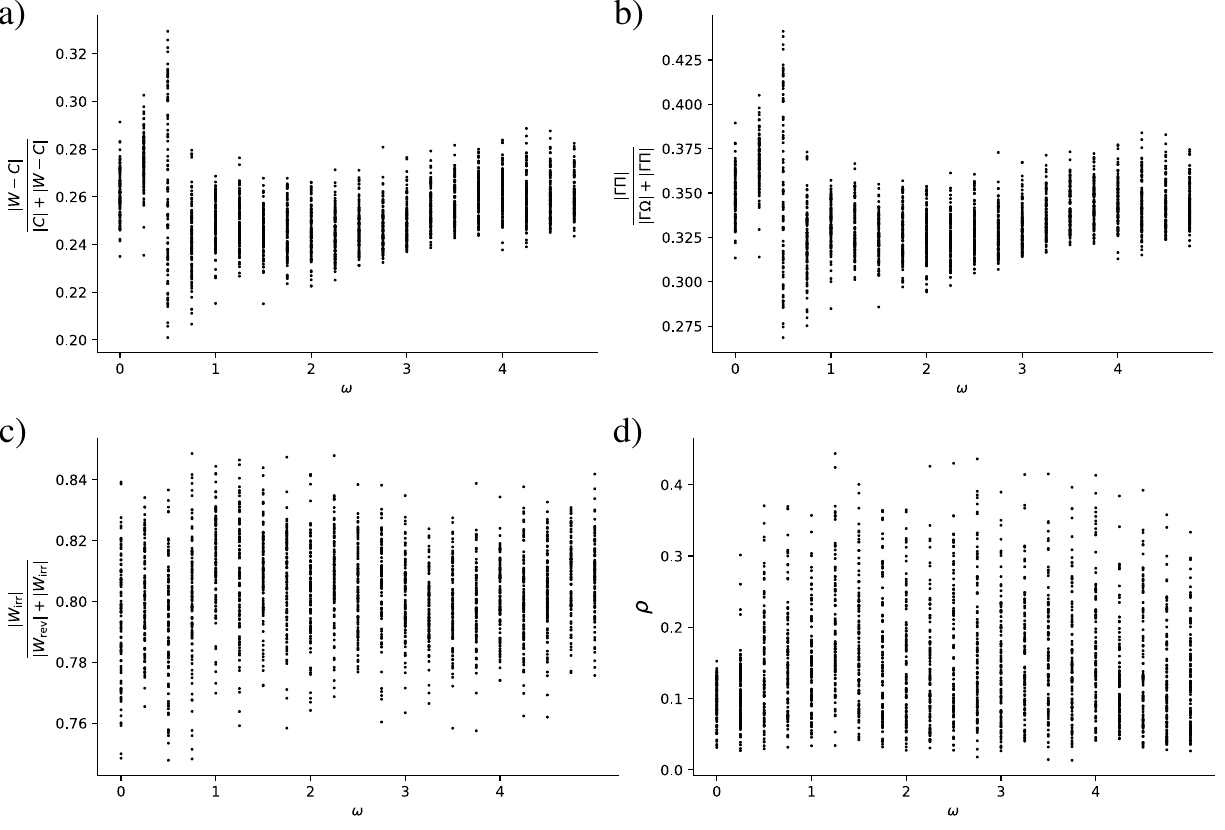}
    \caption{\textbf{Network asymmetry for increasing EPR.} For a range of values $\omega \in [0,5]$ we train an ensemble of 100 RNNs. We then compute the relative size of each component in the decomposition with Eq.~(\ref{eq: network energy}) for $a)$ $(W-C, C)$, $b)$ $(\Gamma \Pi, \Gamma \Omega)$, and $c)$ $(W_{\text{irr}}, W_{\text{rev}})$. In $d)$ we also plot $\varrho$ from the network HHD in Eq.~(\ref{eq: RNN HHD 1}).}
    \label{fig: EPR Asym}
\end{figure*}
We train an ensemble of 100 RNNs to encode the stochastic Hopf oscillator for each value in a range $\omega \in [0,5]$. Fig.~\ref{fig: EPR Asym} shows the relative energy of $a)$ $(W-C, C)$, $b)$ $(\Gamma \Pi, \Gamma \Omega)$, and $c)$ $(W_{\text{irr}}, W_{\text{rev}})$. Panel $d)$ shows the fitted value of $\varrho$ from the network HHD in Eq.~(\ref{eq: RNN HHD 1}). We find that there is no clear relationship between the relative energy in the asymmetric/irreversible component and the EPR, which suggests that irreversible currents are also parametrised in the bias terms, and that the projection matrix $\Gamma$ can induce asymmetries that are not trivially related to the rotational forces in the target vector field. This result contrasts with previous studies suggesting that the EPR is driven directly by network asymmetry \cite{nartallokaluarachchi2024broken}, although this was not in the case of a low-rank system.

\section{Measuring the complexity of dynamics with a universal representation}
\label{app: chaos}
\begin{figure*}
    \centering
    \includegraphics[width=\linewidth]{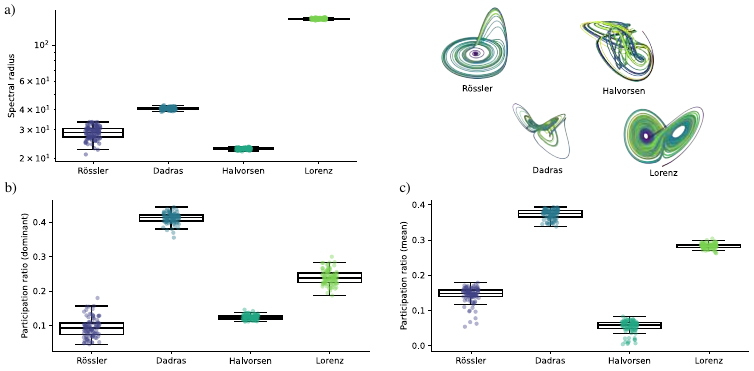}
    \caption{\textbf{A universal representation for nonlinear dynamics.} The DDM framework allows us to obtain a universal representation for nonlinear dynamics. By studying the inferred parameters, we can define measures of `complexity' for the dynamics and their representations. We illustrate this with chaotic attractors, where we compute the $a)$ spectral radius, $b)$ mean participation ratio, and $c)$ participation ratio of the dominant mode, for the connectivity, $W$, of 100 RNNs with 1024 neurons each, trained to encode the Rössler, Dadras, Halvorsen, and Lorenz attractors. We find that the Lorenz attractor has, by far, the highest spectral radius, whilst the Dadras attractor has the highest participation ratio.}
    \label{fig: chaos_complexity}
\end{figure*}
Whilst existing approaches include the calculation of Lyapunov exponents, Kaplan-Yorke embedding dimension, or the entropy rate, there is no consensus on how to measure the `complexity' of a chaotic system \cite{Sprott2010chaos}. Even at a descriptive level it can be difficult to define what is a more or less complex dynamical system. For example, if time is rescaled so that an attractor is traversed at twice the speed, or an attractor is inflated in space, is it of the same complexity?

Whilst this fundamental problem remains, our embedding approach provides a universal representation of a nonlinear system via an RNN. The connectivity structure of the RNN can then be studied to derive a metric for the complexity of the representation, which in turn captures some features of the complexity of the encoded chaotic system. To illustrate this, we consider four familiar 3D chaotic attractors: the Lorenz and Dadras attractors introduced in Sec.~\ref{sec: examples 1}, the \textit{Rössler} attractor,
\begin{align}
    d\begin{pmatrix}
        y_1(t)\\
        y_2(t)\\
        y_3(t)
    \end{pmatrix}& =\begin{pmatrix}
        -y_2-y_3\\
        y_1 + ay_2\\
        b + y_3(y_1-c)
    \end{pmatrix}\;dt, 
\end{align}
with $a=b=0.2$ and $c = 5.7$, and the \textit{Halvorsen} attractor,
\begin{align}
    d\begin{pmatrix}
        y_1(t)\\
        y_2(t)\\
        y_3(t)
    \end{pmatrix}& =\begin{pmatrix}
        -ay_1-4y_2-4y_3-y_2^2\\
        -ay_2-4y_3-4y_1-y_3^2\\
        -ay_3-4y_1-4y_2-y_1^2
    \end{pmatrix}\;dt, 
\end{align}
with $a=1.7$ \cite{Sprott2010chaos}. For each of these systems, we train 100 separate RNNs, with 1024 neurons each, to encode the deterministic drift i.e. we set $\lambda_{\text{diff}} = 0$. With the fitted connectivity matrix, $W$, we then compute the \textit{spectral radius},
\begin{align}
    r(W) & = \max_{i}|\lambda_i|, 
\end{align}
where $\lambda_i$ are the eigenvalues of $W$, and the \textit{participation ratio} (PR),
\begin{align}
    \text{PR}(\mathbf{v}) & = \frac{(\sum_{i} \mathbf{v}_i)^2}{N\sum_{i} \mathbf{v}_i^2},
\end{align}
for each eigenvector $\mathbf{v}$. We can take the mean of the PR over all eigenvectors, or focus on the PR of the dominant mode. The spectral radius captures the maximum asymptotic growth of the system, and the degree of instability for fixed points and periodic trajectories. On the other hand, the participation ratio captures the effective proportion of neurons in the network that are active in encoding a representation. A PR of $1/N$ implies that a single neuron encodes a representation, whilst a PR of $1$ implies the entire network is being used to encode a representation.

Fig.~\ref{fig: chaos_complexity} shows the $a)$ spectral radius, $b)$ PR of the dominant mode, and $c)$ mean PR, for each system and its RNN representations. Whilst there is a lack of `ground truth' for this exploratory experiment, we can notice that the Lorenz system has, by far, the highest spectral radius, suggesting that it has the most instability and asymptotic growth. On the other hand, it is the Dadras attractor which has the highest mean PR and dominant PR, suggesting that it requires the most network complexity to be encoded. This is consistent with the fact that it has the most `scrolls' and thus has non-trivial dynamics in the largest number of directions.

These results are just a first attempt at exploring the complexity of nonlinear systems via the unified and comparable representation offered by DDM, and further progress will require investigating additional network properties beyond those considered here in order to fully capture the richness of the underlying dynamics.

\section{Parameters and details for numerical experiments}
\label{app: params}
Here we provide details on parameters for the numerical simulations. Throughout, we train the perceptrons with \texttt{PyTorch} using the Adam optimiser with learning rate $= 0.001$. We fix $\lambda_{\text{diff}} = 20$. {\color{black} The parameters $\Gamma$ and $W_s$ used standard \texttt{PyTorch} initialisations i.e. they are sampled using a Kaiming uniform initialisation with $\Gamma_{ij}\sim U(-\sqrt{m}, \sqrt{m})$ where $m$ is the number of input dimensions -- thus $m=k$ for $\Gamma$ and $m=n$ for $W_s$. The parameters $\mathbf{b}$ and $I_s$ are initialised as zero vectors. The entries of $B_s$ are initialised using a standard normal distribution i.e. $B_{s,ij} \sim \mathcal{N}(0,1)$.}
\subsection{Nonlinear systems and RNN representations}
In Sec.~\ref{sec: target matching}, we use the following parameters.
\begin{itemize}
    \item \textbf{Van der Pol.}  We take $\sigma = 0.25$. We train an RNN with $N=64$, and 25,000 uniform samples from $[-4,4]^2$ for 30,000 epochs.
    \item \textbf{Lorenz.} We take $\sigma = 0.25$. We train an RNN with $N=512$, and 125,000 uniform samples from $[-25,25]\times [-30,30] \times [0,50]$ with $\lambda_{\text{diff}} = 20$ for 100,000 epochs.
    \item \textbf{Dadras.} We take $\sigma = 0.01$. We train an RNN with $N=1024$, and 150,000 uniform samples from $[-20,20]\times [-12,10] \times [-15,15]$ for 100,000 epochs.
\end{itemize}
\subsection{Switching and cycling attractors}
In Sec.~\ref{sec: attractor design}, we use the following parameters. 
\begin{itemize}
    \item \textbf{Input-driven switching.} We place four minima at $(\pm 3, \pm 3)\in \mathbb{R}^2$, with Gaussian well parameters given by $a_j= 0.125$ and $\nu_j=1$, and noise intensity $\sigma = 0.1$. We train the RNN with $N = 256$ using 25,000 samples from $[-4,4]^2$ for 30,000 epochs.
\end{itemize}
Next, we have the three examples of autonomously cycling diffusions.
\begin{itemize}
    \item \textbf{Three minima.} We place three minima at $\bm{\mu} = \{(1,2), (-1.5,0), (1,-2.25)\}$, with $a_j=0.25$ and $\nu_j=0.75$. The process has $\xi = 0.2$, $\alpha = 1.5$ and $\sigma =0.2$. We train an RNN with $N = 128$ and 25,000 samples from $[-4,4]^2$ for 30,000 epochs.
    \item \textbf{Four minima.} We place four minima at $\bm{\mu} = \{(2,3), (3,-2), (-2,-3), (-1.5,1.5)\}$, with $a_j=0.25$ and $\nu_j=0.75$. The process has $\xi = 0.6$, $\alpha = 2.5$ and $\sigma =0.225$. We train an RNN with $N = 128$ and 25,000 samples from $[-4,4]^2$ for 30,000 epochs.
    \item \textbf{Five minima.} We place five minima at $\bm{\mu} = \{(3.5,4), (0,0), (3,-4), (-3,4), (-3.5,4)\}$, with $a_j=0.25$ and $\nu_j=0.6$. The process has $\xi = 0.6$, $\alpha = 2.5$ and $\sigma =0.225$. We train an RNN with $N = 128$ and 40,000 samples from $[-8,8]^2$ for 40,000 epochs.
\end{itemize}
In Sec.~\ref{sec: network hhd}, we retrain the second layers using 25,000 additional samples and 30,000 epochs to obtain the HHD.

{\color{black}

\subsection{Backpropagation vs. ridge-regression}

In App.~\ref{app: regression} we compare ridge regression with backpropagation at different network sizes.
\begin{itemize}
    \item For the training data, we computed the VDP at 25,000 uniform samples from $[-4,4]^2$.
    \item For the ridge regression, we took $\lambda = 0.01$. We drew $\Gamma_{ij} \sim\mathcal{N}(0,0.1/k)$ and $b_i \sim \mathcal{N}(0,0.01)$.
    \item For the backpropagation, we trained the model for 30,000 epochs.

\end{itemize}

}

\subsection{Non-coplanar examples}
In App.~\ref{app: high dim cycling}, we use the following parameters.
\begin{itemize}
\item \textbf{Pixel images.} We fix minima at the points corresponding to the images, with Gaussian wells specified by $a_j = 7, \nu_j=11$. The process has $\sigma = 0.3$, $\xi = 0.5$, $s=-3.2$, and $\alpha =40$.
\item \textbf{Points on the saddle.} We place points at the positions given by,
\begin{align}
    \bm{\mu}=\begin{pmatrix}
          0.4 & 0.7 & 0.2\\
    1&-0.5&-0.2\\
    -0.2&-0.6&0.4\\
    -0.5&0.2&-0.6
    \end{pmatrix},
\end{align}
and take $r_{\text{max}}=0.4$, $\xi=0.75$ and compute the curl with $\Delta = 10^{-5}$. The Gaussian wells are given by $a_j = 0.5, \nu_j=0.325$ and the process has $\sigma = 0.155$ and $\alpha = 4000$.
\end{itemize}
\subsection{Network asymmetry and entropy production}
In App.~\ref{app: asymm EPR}, we train RNNs with $N=64$ with 25,000 samples in $[-4,4]^2$ for 30,000 epochs and set $\sigma = 0.25$.
\subsection{Measuring complexity of chaotic attractors}
In App.~\ref{app: chaos}, we train RNNs with $N=1024$ with 125,000 samples for 2000 epochs using batches of size 1024. For the Halvorsen attractor, we take the samples in the range $[-15,12]^3$. For the Rössler attractor, we take $[-15,12]\times [-15,12] \times [0,25]$.

\bibliography{Bibliography}
\bibliographystyle{ieeetr}

\end{document}